\documentclass[prc,aps,float,twocolumn,superscriptaddress,showkeys,showpacs,nofootinbib]{revtex4}
\usepackage{amsmath}
\usepackage{amssymb}
\usepackage{amsfonts}
\usepackage{epsfig}
\usepackage{color}
\usepackage{afterpage}
\usepackage[labelfont={normalsize},subrefformat=parens]{subfig}
\usepackage{afterpage}
\newcommand{\old}[1]{{\rule{0cm}{0cm}}}

\newcommand{\oldnote}[1]{{\rule{0cm}{0cm}}}

\newcommand{\antisymV}{\mbox{\boldmath $\mathcal{V}$}}
\newcommand{\rhobold}{\mbox{\boldmath $\rho$}}
\newcommand{\beqn}{\begin{equation}}
\newcommand{\eeqn}{\end{equation}}
\newcommand{\beq}{\begin{equation}}
\newcommand{\eeq}{\end{equation}}
\newcommand{\bea}{\begin{eqnarray}}
\newcommand{\eea}{\end{eqnarray}}
\newcommand{\ba}{\begin{align}}
\newcommand{\ea}{\end{align}}

\newcommand{\btau}{\tau}

\newcommand{\bsig}{ \sigma}

\newcommand{\kvec}{{\bf k}}

\newcommand{\qvec}{{\bf q}}
\newcommand{\pvec}{{\bf p}}

\newcommand{\rvec}{{\bf r}}
\newcommand{\rone}{{\bf r}_1}
\newcommand{\rtwo}{{\bf r}_2}
\newcommand{\rthree}{{\bf r}_3}
\newcommand{\rfour}{{\bf r}_4}

\newcommand{\svec}{{\bf s}}

\newcommand{\Rvec}{{\bf R}}

\newcommand{\WHF}{V_{\rm HF}}

\newcommand{\tr}{{\rm Tr}}

\renewcommand{\vec}[1]{\mathbf{#1}}

\begin{document}

\title{Microscopically-constrained Fock energy density functionals \\
from chiral effective field theory. I. Two-nucleon interactions.}
\author{B. Gebremariam}
\email{gebremar@nscl.msu.edu}
\affiliation{National Superconducting Cyclotron Laboratory,
             1 Cyclotron Laboratory,
             East-Lansing, MI 48824, USA}
\affiliation{Department of Physics and Astronomy,
             Michigan State University, East Lansing, MI 48824, USA}

\author{S. K. Bogner}
\email{bogner@nscl.msu.edu}
\affiliation{National Superconducting Cyclotron Laboratory,
             1 Cyclotron Laboratory,
             East-Lansing, MI 48824, USA}
\affiliation{Department of Physics and Astronomy,
             Michigan State University, East Lansing, MI 48824, USA}
\author{T. Duguet}
\email{thomas.duguet@cea.fr}
\affiliation{National Superconducting Cyclotron Laboratory,
             1 Cyclotron Laboratory,
             East-Lansing, MI 48824, USA}
\affiliation{Department of Physics and Astronomy,
             Michigan State University, East Lansing, MI 48824, USA}
\affiliation{CEA, Centre de Saclay, IRFU/Service de Physique Nucléaire, F-91191 Gif-sur-Yvette, France}
\date{\today}

\begin{abstract}
The density matrix expansion (DME) of Negele and Vautherin is a convenient tool to map finite-range physics associated with vacuum two- and three-nucleon interactions into the form of a Skyme-like energy density functional (EDF) with density-dependent couplings. In this work, we apply the improved formulation of the DME proposed recently in arXiv:0910.4979 by Gebremariam {\it et al.} to the non-local Fock energy obtained from chiral effective field theory (EFT) two-nucleon (NN) interactions at next-to-next-to-leading-order (N$^2$LO).  The structure of the chiral interactions is such that each coupling in the DME Fock functional can be decomposed into a cutoff-dependent coupling {\it constant} arising from zero-range contact interactions and a cutoff-independent coupling {\it function} of the density arising from the universal long-range pion exchanges. This motivates a new microscopically-guided Skyrme phenomenology where the density-dependent couplings associated with the underlying pion-exchange interactions are added to standard empirical Skyrme functionals, and the density-independent Skyrme parameters subsequently refit to data. A Mathematica notebook containing the novel density-dependent couplings is provided.
\end{abstract}



\maketitle


\section{Introduction}
\label{intro}
A longstanding challenge of nuclear theory is to calculate properties of nuclei starting from the vacuum two- and three-nucleon interactions. While impressive progress has been made in extending the limits of ab-initio methods beyond the lightest nuclei~\cite{Pieper:2004qh,Quaglioni:2007qe,Hagen:2008iw}, the nuclear energy density functional (EDF) approach remains the most computationally feasible method for a comprehensive description of medium and heavy nuclei~\cite{bender03b}. Modern parameterizations of empirical Skyrme and Gogny EDFs provide a good description of bulk properties and, to a lesser extent, of certain spectroscopic features of known nuclei. However, the lack a solid microscopic foundation often leads to parameterization-dependent predictions away from known data and makes it difficult to develop systematic improvements.
Fueled by interest in the coming generation of radioactive ion beam facilities, along with studies of astrophysical systems such as neutron stars and supernovae that require controlled extrapolations of nuclear properties in isospin, density, and temperature, there is a large effort currently underway to develop energy functionals with substantially reduced errors and improved predictive power, e.g. see Ref.~\cite{unedf:2007}.

One path forward focuses on empirically improving the analytical forms and fitting procedures of existing phenomenological functionals~\cite{lesinski06a,Lesinski:2007zz,Margueron:2007uf,Niksic:2008vp,carlsson09,Goriely:2009zz}. In the present work, we pursue a complementary approach that relies less on fitting empirical functionals to known data, but rather attempts to constrain the analytical form of the functional and the values of its couplings from many-body perturbation theory (MBPT) and the underlying NN and NNN interactions~\cite{Lesinski:2008cd,Drut:2009ce,Duguet:2009gc,Bogner:2008kj,Kaiser:2003uh,
Kaiser:2009me, Kaiser:2010pp}.

Recent progress in evolving chiral effective field theory (EFT) interactions to lower momentum using renormalization group (RG) methods~\cite{Bogner:2005sn,Bogner:2006vp,Bogner:2006pc,Bogner:2009un,Bogner:2009bt} (see also \cite{Roth:2005ah,Roth:2008km}) plays a significant role in this effort, as the many-body problem formulated in terms of low-momentum interactions is simplified in several key respects. The evolution to low-momentum weakens or largely eliminates non-perturbative
behavior in the two-nucleon sector
arising from strong short-range repulsion and tensor
forces from iterated pion exchanges~\cite{Bogner:2006tw,Bogner:2009bt}. In addition, at lower cutoffs the corresponding three-nucleon interactions
become perturbative and more amenable to approximations such as truncations based on normal-ordering~\cite{Nogga:2004ab,Hagen:2007ew}. When applied to nuclear matter, many-body perturbation theory for the energy appears convergent (at least in the particle-particle channel), with calculations that include all of the NN and most of the NNN second-order contributions,
exhibiting reasonable saturation properties and showing relatively weak dependence on the cutoff \cite{Bogner:2005sn,Bogner:2009un,Hebeler:2009iv}. Moreover, the freedom to vary the order of the input EFT interaction and the cutoff via the RG provides a powerful tool to assess theoretical errors arising from truncations in the Hamiltonian and many-body approximations.

All of these features are favorable ingredients
for the microscopic construction of non-empirical EDFs~\cite{Bogner:2008kj}.
Indeed, Hartree-Fock becomes a reasonable (if not quantitative)
starting point, which
suggests that the theoretical developments and phenomenological
successes of EDF methods for
Coulomb systems may be applicable to the nuclear case for
low-momentum interactions. However, even with these simplifications, perturbative contributions to the energy involve density matrices and propagators folded with finite-range interaction vertices, and are therefore highly non-local in both space and time. In order to make such functionals numerically tractable in heavy open-shell nuclei, it is desirable to develop simplified approximations expressed in terms of the local densities and currents. At lowest order in MBPT (i.e., Hartree-Fock), the density matrix expansion (DME) of Negele and Vautherin~\cite{negele72} can be unambiguously applied to approximate the spatially non-local Fock expression\footnote{We assume local NN interactions since our focus here is on the finite-range pion exchanges, which are local up to an overall cutoff regulator. For non-local interactions, the Hartree contribution is no longer local in space in the sense that it probes the off-diagonal part of the density matrix. } as a generalized Skyrme functional with density-dependent couplings calculated from vacuum interactions. In the present work, we do so on the basis of a chiral NN interaction at next-to-next-to-leading-order (N$^2$LO), while the extension to a chiral NNN interaction at the same order will be discussed in a separate paper.

The non-trivial density dependence of the DME couplings is a consequence of the finite-range of the underlying NN interaction and is controlled by the longest-ranged components. Consequently, the DME offers a path to incorporate physics associated with long-range one- and two-pion exchange interactions into existing Skyrme functionals.  Given the rich spin and isospin structure of such interactions, it is hoped that their inclusion will improve predictive power away from known data and provide microscopic constraints on the isovector structure of nuclear EDFs.

Still, calculations of infinite nuclear matter (INM)~\cite{Bogner:2009un} as well as binding energies and charge radii of doubly-magic nuclei~\cite{Roth:2005ah} demonstrate that it is necessary to go at least to second-order in perturbation theory to resum enough bulk correlations. Furthermore, it is known that while chiral EFT interactions are themselves low-momentum interactions relative to conventional force models, it is still desirable to evolve them to lower-momentum so that HF becomes a reasonable starting point and MBPT is under better control.

However, in the present paper we focus on the lowest-order (i.e., Hartree-Fock) contribution to the energy from the {\it un-evolved} chiral EFT NN interaction. In light of the proceeding remarks, this might appear to be an unrealistic starting point.  This would certainly be the case if our present goal was to develop a fully microscopic and quantitative EDF free from any fitting to data. In the short term, however, we adopt a more pragmatic approach. Our objective in the present approach is to improve existing Skyrme phenomenology by identifying non-trivial density dependencies arising from missing pion physics that can be added to existing Skyrme functionals, which can then be refit to data and implemented in existing codes with minimal modification. The rationale for restricting our attention to the Hartree-Fock energy using un-evolved chiral NN interactions can be summarized as follows:
\begin{itemize}

\item First, it is well-known that the RG evolution to low momentum only modifies the short-distance structure of the inter-nucleon interactions~\cite{Bogner:2003wn,Bogner:2006vp,Bogner:2009bt}. The input chiral NN interactions take the schematic form
\begin{equation}
V^{\text{NN}}_{EFT} = V^{\text{NN}}_{\pi} + V^{\text{NN}}_{ct}(\Lambda)\,,
\end{equation}
where $V^{\text{NN}}_{\pi}$ denotes the finite-range pion-exchange interactions and $V^{\text{NN}}_{ct}(\Lambda)$ denotes scale-dependent zero-range contact terms.
The RG evolution only modifies $V^{\text{NN}}_{ct}$  and leaves the long-distance structure unchanged. Since we are primarily interested in identifying the dominant density dependencies arising from finite-range physics, it is sufficient for our purposes to apply the DME directly to the un-evolved $V^{\text{NN}}_{\pi}$ from the input EFT.

Note that the HF energy arising from $V^{\text{NN}}_{ct}$ bears a strong resemblance to the empirical Skyrme functional (i.e., bilinear products of local densities multiplied by coupling {\it constants}), and therefore does not produce any new density dependencies.

\item Second, a non-trivial extension of the DME is needed to treat non-localities in both space and time that arise in higher orders of perturbation theory. I.e., one must properly account for the presence of energy denominators when designing a DME for 2nd-order MBPT and beyond~\cite{rotival09a}. To date, a satisfactory generalization of the DME has not yet been formulated.

\item Third, even if we were to follow the ad-hoc prescription that consists of replacing the vacuum NN interaction in the Hartree-Fock expression by a Brueckner $G$-matrix (or a perturbative approximation in the case of low-momentum interactions) evaluated at some average energy, the $G$-matrix differs from the NN potential only at short distances.
\end{itemize}

Therefore, while a Hartree-Fock calculation using the un-evolved chiral EFT NN interaction would provide a very poor description of nuclei, the application of the DME to such contributions captures some of the same density dependencies that would arise from the finite-range tail of any in-medium vertex (e.g., a G matrix or a perturbative approximation thereof) that sums ladder diagrams in a more sophisticated many-body treatment.  Once a satisfactory generalization of the DME is developed to handle spatial and temporal non-locality on the same footing, non-localities arising from in-medium propagation can be mapped into the density-dependent Skyrme couplings as well\footnote{It remains to be seen if such a generalization will necessitate the introduction of orbital-dependent terms into the EDF.}.

\begin{table}
\begin{center}
\begin{tabular}{ll}
\hline \hline \\
EDF &   Energy density functional \\
DME &   Density matrix expansion\\
PSA &   Phase space averaging\\
NV   & Negele and Vautherin\\
OBDM &  One-body density matrix\\
INM &   Infinite nuclear matter\\
MBPT & Many-body perturbation theory \\
HF & Hartree-Fock\\
EFT & Effective field theory \\
& \\
\hline \hline
\end{tabular}
\caption{List of acronyms repeatedly used in the text.}
\label{table:acronym}
\end{center}
\end{table}

The rest of the paper is organized as follows. In Sec.~\ref{sec:HF} we derive the Hartree-Fock energy for even-even nuclei, which serves as the starting point for the DME. Section~\ref{sec:DME} reviews the improved PSA-DME of Ref.~\cite{Gebremariam:2009ff} that is used in the present work. Master formulas and skeleton expressions for the resulting DME couplings obtained from the Fock energy are given in Sec.~\ref{sec:Couplings}. Results for various density-dependent couplings are discussed in Sec.~\ref{sec:Results} and conclusions are given in Sec.~\ref{sec:Conclusions}.  Various technical details and lengthy expressions are given in the appendices. The explicit forms for the chiral EFT NN finite range and contact interactions are given in Appendices~\ref{appendix:NNpot} and ~\ref{appendix:contacts}, and the exchange interaction is given in Appendix~\ref{appendix:exchangeNN}.  The PSA-DME is reviewed in Appendix~\ref{appendix:psadme}. Formulas to construct the single particle fields obtained from the density-dependent DME couplings are given in Appendix~\ref{appendix:spfields}, and the couplings that arise from performing the DME on the Hartree energy are collected in Appendix~\ref{appendix:hartree}. Finally, detailed expressions of the DME couplings are provided in a companion Mathematica notebook, and are also shown in Appendix~\ref{appendix:couplings}.

\section{HF energy for even-even nuclei}
\label{sec:HF}

Before applying the DME to the Fock contribution from the chiral NN interaction, it is useful to provide a detailed expression of the Hartree-Fock potential energy, $\WHF$. We restrict our attention to the ground states of even-even nuclei throughout this paper, with the consequence that certain contributions to the energy are zero due to the intrinsic time-reversal invariance of such states. Note however that the PSA-DME of Ref.~\cite{Gebremariam:2009ff} provides a natural framework to extend the approach to states that break time-reversal symmetry, i.e. ground states of odd-even or odd-odd nuclei, see Appendix~\ref{appendix:psadme}. For a general (possibly non-local)
two-nucleon potential $V^{\text{NN}}$, $\WHF$ is defined in terms of occupied self-consistent HF orbitals as
\bea
\label{eq:HFdef}
 \WHF &=&\frac{1}{2}
\sum_{i j}^{A} \langle i j | V^{\text{NN}}(1-P_{12})|i j\rangle \nonumber \\
&\equiv& \frac{1}{2}\sum_{i j}^{A} \langle i j | \mathcal{V}^{\text{NN}}|i
j\rangle  \;.
\eea
The antisymmetrized interaction $\mathcal{V}\equiv  V^{\text{NN}} (1-P_{12})$  has
been introduced, with $P_{12}$ equal to
the product of spin, isospin and space two-body exchange operators
$P_{12} \equiv P_{\sigma}P_{\tau}P_r$, where
\beqn
P_{\sigma} \equiv \frac{1}{2}(1 + \bsig_1\!\cdot\!\bsig_2)\quad {\rm and}\quad P_{\tau} \equiv\frac{1}{2}(1 + \btau_1\!\cdot\!\btau_2)\,.
\eeqn
By making repeated use of the completeness relation
\beqn
 \openone =  \sum_{\sigma\tau}\int\!d{\bf r}\,|{\bf r}\sigma\tau\rangle\langle
{\bf r}\sigma\tau|
  \;,
\eeqn
and the definition of the HF density matrix
\beqn
  \label{eq:densitymat}
  \rho(\rthree\sigma_3\tau_3,\rone\sigma_1\tau_1) \equiv
  \sum_{i}^{A}\phi^*_{i}(\rone\sigma_1\tau_1)\phi_{i}(\rthree\sigma_3\tau_3)
  \;,
\eeqn
Eq.~(\ref{eq:HFdef}) can be written as
\bea
  \label{eq:HF_DM1}
  \WHF &=&\frac{1}{2} \sum_{\{\sigma \tau\}}
  \!\int\! \prod_{i=1}^{4} d{\bf r}_{i}\,
  \langle \rone\sigma_1\tau_1\rtwo\sigma_2\tau_2|\mathcal{V}^{\text{NN}}
  |\rthree\sigma_3\tau_3\rfour\sigma_4\tau_4\rangle
  \nonumber
  \\
  &&\hspace{.5in}\null\times
  \rho(\rthree\sigma_3\tau_3,\rone\sigma_1\tau_1)
  \rho(\rfour\sigma_4\tau_4,\rtwo\sigma_2\tau_4)\nonumber
 \\ [.15in]
  &=&\frac{1}{2}\tr_{1}\tr_{2}
  \!\int\! \prod_{i=1}^{4}d{\bf r}_i\, \langle\rone\rtwo| \antisymV^{\text{NN}}_{1\otimes 2} |\rthree\rfour\rangle\nonumber\\
  &&\hspace{.9in}\null\times \rhobold^{(1)}(\rthree,\rone)\rhobold^{(2)}(\rfour,\rtwo),
\eea
where a matrix notation in spin and isospin spaces is used
in the second equation and the traces denote
summations over spin and isospin indices
for ``particle 1" and ``particle 2''. Hereafter we drop the
indices on  $\antisymV^{\text{NN}}$ and $\rhobold$ indicating
which space they act in as it will be clear from the context. Switching to relative/center-of-mass (COM) coordinates and noting that the free-space two-nucleon potential is diagonal in the COM coordinate,
the Hartree-Fock expression becomes
\bea
\WHF&=&\frac{1}{2}\tr_{1}\tr_{2}
  \!\int\! d \Rvec\, d\rvec\,d\rvec'\langle\rvec'| \antisymV^{\text{NN}} |\rvec\rangle \nonumber\\
&&\hspace{.3in}\null\times\rhobold(\Rvec+\rvec/2,\Rvec+\rvec'/2)\nonumber\\
&&\hspace{.6in}\null\times\rhobold(\Rvec-\rvec/2,\Rvec-\rvec'/2),
 \label{eq:WHF}
\eea
where the antisymmetrized coordinate space interaction is given by the Fourier transform
\bea
\hspace{-.4in}\langle \rvec|\antisymV^{\text{NN}}|\rvec'\rangle &=& \int\frac{d\pvec\,d\pvec'}{(2\pi)^6}e^{i\pvec\cdot\rvec'}e^{-i\pvec'\cdot\rvec}
\biggl(\langle\pvec'|\mbox{\boldmath $V$}^{\text{NN}}|\pvec\rangle \nonumber\\
&&\hspace{.9in}-\,\, \langle\pvec'|\mbox{\boldmath $V$}^{\text{NN}}P^{\sigma\tau}|-\pvec\rangle\biggr)\,,
\eea
where $\pvec'$ and $\pvec$ are the ``incoming" and ``outgoing" relative momenta and $\langle\pvec|\mbox{\boldmath $V$}^{\text{NN}}|\pvec'\rangle$ is understood to be an operator with respect to spin/isospin quantum numbers and a matrix element with respect to momentum. For chiral NN interactions through
N$^2$LO,  the explicit spin/isospin structure can be expressed as
\begin{widetext}
\begin{eqnarray}
 \langle \pvec' |\mbox{\boldmath $V$}^{\text{NN}}|\pvec\rangle&\equiv&\, \bigl[V_C \, + \,\tau_1  \cdot \tau_2 W_C \bigr]\, + \, \bigl[ \,
\, V_S \, + \,\tau_1 \cdot \tau_2 \,W_S\, \bigr]\,\vec{ \sigma}_1
\cdot\vec{\sigma}_2 \, + \, \bigl[\, V_T \, + \,\tau_1 \cdot \tau_2
\,W_T \, \bigr] \, \vec{\sigma}_1 \cdot \vec{q} \,  \vec{\sigma}_2
\cdot \vec{q } \, \nonumber \\
&+&\,\bigl[\widetilde{V}_C \,+\, \tau_1  \cdot \tau_2 \widetilde{W}_C \bigr]\,
+ \, \bigl[ \,
\, \widetilde{V}_S \, + \,\tau_1 \cdot \tau_2 \,\widetilde{W}_S\, \bigr]\,\vec{ \sigma}_1
\cdot\vec{\sigma}_2 \, + \, \bigl[\, \widetilde{V}_T \, + \,\tau_1 \cdot \tau_2
\,\widetilde{W}_T \, \bigr] \, \vec{\sigma}_1 \cdot \vec{k} \,  \vec{\sigma}_2
\cdot \vec{k } \, \nonumber \\
&+& \, \bigl[\, V_{LS} \, + \,\tau_1 \cdot \tau_2 \,W_{LS} \,
\bigr] \, \frac{i}{2} \,\bigl(\, \vec{\sigma}_1 \,+ \, \vec{\sigma}_2 \, \bigr)
\cdot \bigl(\, \vec{q} \,\times  \vec{k} \, \bigr) \,,
\label{eq:EFTdecomp}
\end{eqnarray}
where the two-body form factors $\{V_C, V_S,\ldots\}$ are functions of the momentum transfer $\qvec = \pvec'-\pvec$ and $\{\widetilde{V}_C,\widetilde{V}_S,\ldots\}$ are functions of $\kvec = (\pvec + \pvec')/2$.  Explicit expressions for the momentum-space form factors are given in Appendix~\ref{appendix:NNpot}. The $\kvec$-dependent terms arise entirely from the zero-range contact terms in the chiral EFT potential, while the $\qvec$-dependent terms receive contributions from both finite-range pion exchanges and zero-range contact terms. Decomposing $\langle\pvec'|\mbox{\boldmath $V$}^{\text{NN}}P^{\sigma\tau}|-\pvec\rangle$ analogously to Eq.~\ref{eq:EFTdecomp}, one has
\bea
\langle \pvec' |\mbox{\boldmath $V$}^{\text{NN}}P^{\sigma\tau}|-\pvec\rangle&\equiv&\, \bigl[V_C^x \, + \,\tau_1  \cdot \tau_2 W_C^x \bigr]\, + \, \bigl[ \,
\, V_S^{x} \, + \,\tau_1 \cdot \tau_2 \,W_S^{x}\, \bigr]\,\vec{ \sigma}_1
\cdot\vec{\sigma}_2 \, + \, \bigl[\, V_T^x \, + \,\tau_1 \cdot \tau_2
\,W_T^x \, \bigr] \, \vec{\sigma}_1 \cdot \vec{k} \,  \vec{\sigma}_2
\cdot \vec{k } \, \nonumber \\
&+&\,\bigl[\widetilde{V}_C^x \,+\, \tau_1  \cdot \tau_2 \widetilde{W}_C^x \bigr]\,
+ \, \bigl[ \,
\, \widetilde{V}_S^x \, + \,\tau_1 \cdot \tau_2 \,\widetilde{W}_S^x\, \bigr]\,\vec{ \sigma}_1
\cdot\vec{\sigma}_2 \, + \, \bigl[\, \widetilde{V}_T^x \, + \,\tau_1 \cdot \tau_2
\,\widetilde{W}_T^x \, \bigr] \, \vec{\sigma}_1 \cdot \vec{q} \,  \vec{\sigma}_2
\cdot \vec{q } \, \nonumber \\
&+& \, \bigl[\, V_{LS}^x \, + \,\tau_1 \cdot \tau_2 \,W_{LS}^x \,
\bigr] \, \frac{i}{2} \,\bigl(\, \vec{\sigma}_1 \,+ \, \vec{\sigma}_2 \, \bigr)
\cdot \bigl(\, \vec{k} \,\times  \vec{q} \, \bigr) \,,
\label{eq:xEFTdecomp}
\eea
\end{widetext}
where the exchange force form factors $\{V_C^x, V_S^x,\ldots\}$ are functions of $\kvec$ whereas $\{\widetilde{V}_C^x,\widetilde{V}_S^x,\ldots\}$ are functions of $\qvec$. The exchange form factors can be expressed as linear combinations of the direct interaction, and are given in Appendix~\ref{appendix:exchangeNN}. 

Momentum space NN potentials that depend only on $\qvec$ or $\kvec$ are sometimes referred to as quasi-local, as their coordinate space representation is diagonal such that the anti-symmetrized interaction in Eq.~\ref{eq:HF_DM1} is schematically given by
\begin{equation}
\langle \rvec|\antisymV^{\text{NN}}|\rvec'\rangle = \delta(\rvec-\rvec')\mbox{\boldmath $V$}(r,\nabla) - \delta(\rvec+\rvec')\mbox{\boldmath $V$}(r,\nabla)P^{\sigma\tau}\,.
\end{equation}
For the chiral potentials considered here, the $\mbox{\boldmath $V$}(r,\nabla)$ can have terms that are linear (via the spin-orbit interaction) or quadratic (via the contact interactions) in $\nabla$. In order to preserve the quasi-local nature of the potential, we neglect the ultra-violet (UV) regulator that multiplies each term in Eq.~\ref{eq:EFTdecomp}. The most commonly used regulators give a non-local interaction since $\langle\pvec'|V^{\text{NN}}|\pvec\rangle$ in Eq.~\ref{eq:EFTdecomp} is replaced by $f(p'/\Lambda)\langle\pvec'|V^{\text{NN}}|\pvec\rangle f(p/\Lambda)$, where $f(p/\Lambda)\rightarrow 0$ for $p\gg\Lambda$ and $f(p/\Lambda)\approx 1$ for $p\ll\Lambda$. While it is possible to use a regulator that suppresses large momentum {\it transfers} instead of large relative momenta, it is reasonable to neglect the regulator since we work at the HF level with a local fermi momentum $k_F\ll\Lambda$. Note that the UV cutoff $\Lambda$ is between $\sim 2.5\!-\!3.0$ fm$^{-1}$ in most implementations of the N$^2$LO NN interactions.

Given that we are working with a quasi-local interaction, it is convenient to treat Hartree and Fock contributions to Eq.~\ref{eq:WHF} separately. To do so, we first expand the $\rhobold$ matrix on Pauli spin and isospin matrices
\bea
\rhobold(\rvec_1,\rvec_2) &=& \frac{1}{4}[\rho_0(\rvec_1,\rvec_2) + \rho_1(\rvec_1,\rvec_2)\tau_z \nonumber\\
                     &&  +\,\, \vec{S}_0(\rvec_1,\rvec_2)\!\cdot\!\vec{\sigma}
           +\vec{S}_1(\rvec_1,\rvec_2)\!\cdot\!\vec{\sigma}\tau_z]
	   \;,
\label{eq:DMdecomp}
\eea
where the usual scalar-isoscalar, scalar-isovector, vector-isoscalar, and vector-isovector components are obtained by taking the relevant traces,
\bea
 \rho_0(\rvec_1,\rvec_2) &\equiv&
 \tr_{\sigma\tau}[\rhobold(\rvec_1,\rvec_2)]
 \;,
 \\
 \rho_1(\rvec_1,\rvec_2) &\equiv&
 \tr_{\sigma\tau}[\rhobold(\rvec_1,\rvec_2)\tau_z]
 \;,
 \\
 \vec{S}_0(\rvec_1,\rvec_2) &\equiv&
 \tr_{\sigma\tau}[\rhobold(\rvec_1,\rvec_2)\vec{\sigma}]
 \;,
 \\
  \vec{S}_1(\rvec_1,\rvec_2) &\equiv&
 \tr_{\sigma\tau}[\rhobold(\rvec_1,\rvec_2)\vec{\sigma}\tau_z]
\;.
   \label{eq:densitymatrix}
\eea
Starting from these non-local densities, it is useful to define the following set of local densities ($t=0,1$)
\begin{eqnarray}
\rho_{t}(\vec{R})  &\equiv& \rho_t (\vec{r}_1, \vec{r}_2) |_{\vec{r}_1=\vec{r}_2=\vec{R}} \,\,\, , \\
\tau_{t}(\vec{R}) &\equiv&  \nabla_1 \cdot \nabla_2 \, \rho_t (\vec{r}_1, \vec{r}_2) |_{\vec{r}_1=\vec{r}_2=\vec{R}} \,\,\,,\\
\vec{J}_{t}(\vec{R}) &\equiv& - \frac{i}{2} (\vec{\nabla}_1 - \vec{\nabla}_2)\times \; \vec{S}_{t} (\vec{r}_1,\vec{r}_2) |_{\vec{r}_1=\vec{r}_2=\vec{R}} \,\,\, ,
\end{eqnarray}
which correspond to the matter density, the kinetic density and the spin-orbit current density, respectively.

Inserting  Eqs.~\ref{eq:EFTdecomp} and \ref{eq:xEFTdecomp} into Eq.~\ref{eq:WHF}, evaluating the spin/isospin traces and dropping terms that vanish in even-even nuclei (e.g., terms involving the local part of the spin density ${\bf S}(\rvec,\rvec)$) finally gives
\begin{widetext}
\beq
V_H = \frac{1}{2}\sum_{t=0,1}\int d\Rvec d\rvec\Bigl[\rho_t(\Rvec+\rvec/2)\rho_t(\Rvec-\rvec/2)\,\Gamma^t_C(\rvec) + \rvec\!\cdot\!{\bf J}_t(\Rvec+\rvec/2)\rho_t(\Rvec-\rvec/2)\,\Gamma^t_{LS}(\rvec)\Bigr] \,,\label{eq:finalH}
\eeq
and
\bea
V_F &=&
 - \frac{1}{2}\sum_{t=0,1}\int d\Rvec d\rvec \Bigl[\rho_t^2(\Rvec+\rvec/2,\Rvec-\rvec/2)\,\Gamma^{xt}_C(\rvec) - {\bf S}_t^2(\Rvec+\rvec/2,\Rvec-\rvec/2)\,\Gamma^{xt}_S(\rvec) \label{eq:finalF} \\
 &&\qquad\qquad\quad\quad +\,\,\, S^{\alpha}_t(\Rvec+\rvec/2,\Rvec-\rvec/2)S^{\beta}_t(\Rvec+\rvec/2,\Rvec-\rvec/2)\,\nabla_{\alpha}\nabla_{\beta}\Gamma^{xt}_T(\rvec)\Bigr] \nonumber\\
 &&+ i \sum_{t=0,1}\int d\rvec_1d\rvec_2\, \Gamma^{xt}_{LS}(\rvec)\,{\bf S}_t(\rvec_2,\rvec_1)\cdot\bigl(\rvec\times\nabla_1\bigr)\rho_t(\rvec_1,\rvec_2)\nonumber \,,
\eea
\end{widetext}
where $V_H$ and $V_F$ denote the direct (Hartree) and exchange (Fock) contributions, respectively. The $\Gamma$ vertices, which in fact only depend on the norm of $\rvec$, are defined as
\bea
\Gamma^t_i(\rvec) &\equiv& V_i(\rvec) - \widetilde{V}^x_i(\rvec) \quad\,\,\,\, t = 0 \,\, ,\\
                                    &\equiv& W_i(\rvec) -\widetilde{W}^x_i(\rvec) \quad t = 1 \,\, ,\nonumber
\eea
\bea
\Gamma^{xt}_i(\rvec) &\equiv& V^x_i(\rvec) - \widetilde{V}_i(\rvec) \quad\,\,\,\, t = 0 \,\, ,\\
                                    &\equiv& W^x_i(\rvec) -\widetilde{W}_i(\rvec) \quad t = 1 \,\, ,\nonumber
\eea
for $i \in \{C,S,T,LS\}$, where the coordinate-space interactions are given by, e.g.,
\bea
V_i(\rvec) &\equiv& \int \frac{d\qvec}{(2\pi)^3} e^{i\qvec\rvec} \,V_i(\qvec)\quad \rm{for\,\,i} = C, S, T, \\
&\equiv& \frac{i}{r^2}\int \frac{d\qvec}{(2\pi)^3} e^{i\qvec\rvec} \,(\qvec\!\cdot\!\rvec)\, V_i(\qvec) \,\, \rm{for}\,\,i = LS\,.
\eea

As discussed in Ref.~\cite{Gebremariam:2009ff}, our primary focus is to apply the DME to the exchange (Fock) part of the HF energy while treating the Hartree term exactly. Indeed, it was realized long ago, starting with the early works by Negele and Vautherin~\cite{negele72,negele75}, that
\begin{itemize}
\item[(i)] Treating the Hartree contribution exactly provides a better reproduction of the density fluctuations and the energy produced from an exact HF calculation~\cite{negele75}.
\item[(ii)] Restricting the DME to the exchange contribution significantly reduces the self-consistent propagation of errors~\cite{negele75,sprung75}.
\item[(iii)] Treating the Hartree contribution exactly generates no additional complexity in the numerical solutions of the resulting self-consistent HF equations~\cite{negele75} compared
to applying the DME to both Hartree and Fock terms.
\end{itemize}
Nevertheless, it is possible to apply the DME to the Hartree terms so that the complete Hartree-Fock contribution is mapped into a local EDF. For completeness, the DME couplings arising from the Hartree contributions are also collected in Appendix~\ref{appendix:hartree}. However, owing to the deficiencies of the DME at reproducing such contributions, it is strongly advised to treat the finite range Hartree terms exactly in actual self-consistent EDF calculations~\cite{stoitsov09a}. See, however, Ref.~\cite{Dobaczewski:2010qp} where an accurate local EDF approximation for the Hartree energy for the Gogny force is obtained by performing a simple Taylor series expansion.
\section{Density Matrix Expansion}
\label{sec:DME}
Long ago, Negele and Vautherin formulated the density matrix expansion to establish a theoretical connection between the phenomenological Skyrme energy functional, which is written as sums over bilinear products of local densities and currents, and microscopic Hartree-Fock expressions involving the non-local density matrix and finite range NN interaction \cite{negele72,negele75}.
The central idea is to factorize the non-locality of the one-body density matrix (OBDM) by expanding it into a finite sum of terms that are separable in relative and center of mass coordinates. Adopting notations similar to those introduced in
Ref.~\cite{doba03b}, one writes
\begin{eqnarray}
\rho_{t} (\vec{r}_1, \vec{r}_2)  &\approx&  \sum^{n_{\text{max}}}_{n=0} \Pi^{\rho}_n (k  r) \,\, {\cal P}_n (\vec{R}) \,\,\, , \label{approxscalar}\\
\vec{S}_t (\vec{r}_1, \vec{r}_2) &\approx& \sum^{m_{\text{max}}}_{m=0} \Pi^{\vec{s}}_m (k r) \,\, {\cal Q}_m (\vec{R}) \,\,\, ,\label{approxvector}
\end{eqnarray}
where $k$ is a yet-to-be-specified momentum that sets the scale for the decay in the off-diagonal direction, whereas $\Pi^f_n (k r)$ denote the so-called
$\Pi-$functions that also remain to be specified.\footnote{In the PSA-DME, the $\Pi$-functions are the same for the scalar and vector parts,  $\Pi^{\rho}_n = \Pi^{{\bf s}}_n$. See Appendix~\ref{appendix:psadme} for details.} Functions $\{{\cal P}_n
(\vec{R}), {\cal Q}_m (\vec{R})\}$ denote various local densities and their gradients $\{\rho_t(\vec{R}), \tau_{t}(\vec{R}), \vec{J}_{t}(\vec{R}), \vec{\nabla} \rho_t (\vec{R}),
\Delta \rho_t(\vec{R})\}$.

The benefit of expansion~\ref{approxscalar}-\ref{approxvector} is to approximate the non-local Fock energy (Eq.~\ref{eq:finalF}) as a bilinear local Skyrme-like EDF of the form (for time-reversal invariant systems)
\begin{widetext}
\bea
V_F &\approx& \sum_{t=0,1}\int d\Rvec \biggl[ C_t^{\rho\rho}\rho_t^2 \,+\, C_t^{\rho\tau}\rho_t\tau_t \,+\, C_t^{\rho\Delta\rho}\rho_t\Delta\rho_t \,+\, C_t^{\nabla\rho\nabla\rho}\bigl(\nabla\rho_t\bigr)^2\,+\,C_t^{J\nabla\rho}\vec{J}\!\cdot\!\vec{\nabla}\rho_t\,+\,C_t^{JJ}\vec{J}_t^2\biggr]\,,
\label{eq:EDF}
\eea
\end{widetext}
where the densities depend locally on $\Rvec$, while the couplings that are microscopically derived from the vacuum NN interaction depend on the arbitrary momentum scale $k$. In the present work, we adopt the usual LDA choice where $k$ is chosen to be the local Fermi momentum related to the isoscalar density through
\begin{equation}
k \equiv k_F(\vec{R}) = \biggl(\frac{3\pi^2}{2}\rho_0(\vec{R})\biggr)^{1/3}\,\, , \label{LDA}
\end{equation}
although other choices are possible that include additional $\tau$- and $\Delta\rho$-dependencies~\cite{campi77}. The DME couplings are therefore density-dependent (or equivalently $\Rvec$-dependent) and are given by integrals of the finite-range NN interaction over various combinations of the $\Pi$-functions, e.g.
\begin{equation}
C_t^{\rho\tau}(\Rvec) \sim \int dr\,r^2\,\Pi_0^{\rho}(k_Fr)\,\Pi_2^{\rho}(k_Fr)\, \Gamma_c^{xt}(r)\,,
\end{equation}
where the $\vec{R}$-dependence of $k_F$ is suppressed for brevity. In Eq.~\ref{eq:EDF}, only pseudovector contributions $\vec{J}_t^2$ to the more complete ``tensor terms''~\cite{Bender:2009ty} have been kept for simplicity. While this is exact in spherical nuclei, it is only approximate in nuclei that break rotational invariance. However, pseudotensor contributions have recently been shown~\cite{Bender:2009ty} to be systematically two orders
of magnitude smaller than vector ones in axially deformed nuclei, which justifies the common practice of neglecting the former for the purpose of calculating binding energies in situations where Galilean invariance is not broken.

Several DME variants have been developed in the past~\cite{Gebremariam:2009ff,negele72,campi77,meyer86,soubbotin99}. They mainly differ regarding (i) the choice of the momentum scale $k$, (ii) the path followed to obtain actual expressions of the $\Pi-$functions (see below) and (iii) the set of local densities that occur in the
expansion. As discussed in Ref.~\cite{Gebremariam:2009ff}, all of these variants give reasonably accurate descriptions of the Fock energy contributions that probe the scalar part of the OBDM.

However, for the (spin) vector part of the OBDM, which is relevant for approximating Fock contributions in spin-unsaturated nuclei where at least one pair of spin-orbit partners is only partially filled, the situation is very different. In Ref.~\cite{Gebremariam:2009ff}, it was shown that the poor accuracy of the original DME of Negele and Vautherin (NV-DME) for the vector part of the OBDM can be dramatically improved by using phase space averaging (PSA) techniques and by accounting for the anisotropy of the local momentum distribution that is a generic feature of {\it finite} Fermi systems~\cite{bulgac96}. The anisotropy is especially pronounced in the nuclear surface region where the vector part of the OBDM is sharply peaked, and should be accounted for if the DME is to provide an accurate description of spin-unsaturated nuclei that constitute the majority of nuclei.

In this paper we present results for both the standard NV-DME~\cite{negele72} and a simplified variant of the recently developed PSA-DME~\cite{Gebremariam:2009ff}.  For both versions, we follow common practice and truncate the DME to $n_{\text{max}}=2$ in Eq.~\ref{approxscalar} for the scalar part

\begin{widetext}
\begin{eqnarray}
\rho_t (\vec{R} +  \frac{\vec{r}}{2},\vec{R} -
\frac{\vec{r}}{2} )& \simeq & \Pi^{\rho}_{0} (k_F r) \, \rho_t
(\vec{R})+\frac{r^2}{6}\Pi^{\rho}_{2} (k_F r)
\biggl[\frac{1}{4} \Delta \rho_t (\vec{R})- \tau_{t} (\vec{R}) +
\frac{3}{5} k^{2}_F \rho_{t} (\vec{R})\biggr]\,, \label{eq:DMEscalar}
\end{eqnarray}
\end{widetext}
and to $m_{\text{max}}=1$ in Eq.~\ref{approxvector} for the vector part
\bea
\vec{S}_t(\vec{R}+\frac{\vec{r}}{2}, \vec{R}-\frac{\vec{r}}{2}) &\simeq& -\frac{i}{2}\Pi^{\vec{s}}_1 (k_F r)\,\rvec\,\times \vec{J}_t(\Rvec)\,,
\label{eq:DMEvector}
\eea
with the additional feature that the $n=1$ ($m=0$) contribution to the scalar (vector) part of the OBDM is zero in time-reversal invariant systems. The NV-DME $\Pi$-functions are given by,
\bea
\label{eq:NVpi}
\Pi^{\rho}_0(k_Fr) \,&=&\, 3\,\frac{j_1(k_F(\Rvec)r)}{k_F(\Rvec)r}\,, \\
\Pi^{\rho}_2(k_Fr) \,&=&\, 105\,\frac{j_3(k_F(\Rvec)r)}{(k_F(\Rvec)r)^3} \,, \\
\Pi^{{\bf s}}_1(k_Fr) \,&=&\, j_0(k_F(\Rvec)r)\,,
\label{eq:pi0pi2}
\eea
whereas the PSA-DME vector $\Pi$-functions are given by
\begin{equation}
\Pi^{\rho}_0(k_Fr) = \Pi^{\rho}_2(k_Fr)= \Pi^{\svec}_1(k_Fr)\,=\, 3\,\frac{j_1(\widetilde{k}_F(\Rvec)r)}{\widetilde{k}_F(\Rvec)r}\,,
\label{eq:pi1}
\end{equation}
with
\bea
\widetilde{k}_F(\Rvec) \equiv \biggl(\frac{2 + 2P_2(\Rvec)}{2-P_2(\Rvec)}\biggr)^{1/3}k_F(\Rvec)\,.
\label{eq:kFdeformed}
\eea
The function $P_2(\Rvec)$ denotes the quadrupole anisotropy of the local momentum distribution, which can be calculated from the Husimi phase-space distribution\footnote{The Husimi distribution $H(\rvec,\vec{p})$ is a positive-definite generalization of the Wigner $f(\rvec,\vec{p})$ function, see Ref.~\cite{husimi40}.}
\begin{eqnarray}
P_2(\vec{r})\,&\equiv&\,\frac{\int d\vec{p} \bigl[3(\vec{e}_r \cdot
\vec{p})^2 -\vec{p}^2 \bigr] H(\vec{r},\vec{p})}{\int d \vec{p} \,
\vec{p}^2 H(\vec{r},\vec{p})} \nonumber \\
&\simeq& \biggl[\frac{3}{\tau_0(\vec{r})} \sum_{i=1}^{A} |(\vec{e}_r \cdot \vec{\nabla})
\varphi_i (\vec{r} )|^2 \,  -1\biggr] \,, \label{eqn:P_2definition}
\end{eqnarray}
where  $\varphi_i(\rvec)$ denotes an occupied HF single particle spinor with components $\varphi_i(\rvec\sigma\tau)$.

As noted in Ref.~\cite{Gebremariam:2009ff} and discussed in Appendix~\ref{appendix:psadme}, there is a simplified PSA approximation that uses the phase space of infinite nuclear matter and amounts to setting   $P_2(\Rvec)=0$ in Eq.~\ref{eq:pi1}. From a practical perspective, the simplified PSA-DME provides a convenient compromise as it
gives substantial improvements over the NV-DME in describing the vector part of the OBDM while avoiding the complicated single-particle fields that arise from the $P_2(\Rvec)$ dependence in the full PSA-DME. This is the version actually used in the present work.

\section{DME Couplings}
\label{sec:Couplings}
\subsection{Separating long- and short-distance contributions}
Before evaluating the DME couplings, it is convenient to notice that a clean separation between long- and short-distance physics can be made due to the generic structure of the EFT interactions. Schematically, the EFT potential can be written as
\bea
V^{\text{NN}} &=& V^{\text{NN}}_{1\pi} + V^{\text{NN}}_{2\pi} +\ldots+ V^{\text{NN}}_{ct}(\Lambda)\,,
\eea
where $V^{\text{NN}}_{1\pi}$ and $V^{\text{NN}}_{2\pi}$ are finite-range one- and two-pion exchange interactions dictated by the spontaneously broken chiral symmetry of QCD, and $V^{\text{NN}}_{ct}$ denotes scale-dependent contact terms that encode the effects of integrated out degrees of freedom (heavier meson exchanges, high-momentum two-nucleon states, etc.) on low energy physics. Through N$^2$LO, the contact interaction takes the form
\bea
\label{Vcon}
\langle \pvec|V^{\text{NN}}_{ct}|\pvec'\rangle &=& C_S  + C_T \, \vec{\sigma}_1 \cdot  \vec{\sigma}_2 \,+\, C_1 \, \vec{q}\,^2 + C_2 \, \vec{k}^2 \nonumber \\
&+&\,
( C_3 \, \vec{q}\,^2 + C_4 \, \vec{k}^2 ) ( \vec{\sigma}_1 \cdot \vec{\sigma}_2)\nonumber\\
&+& iC_5\, \frac{1}{2} \, ( \vec{\sigma}_1 + \vec{\sigma}_2) \cdot ( \vec{q} \times
\vec{k})+ C_6 \, (\vec{q}\cdot \vec{\sigma}_1 )(\vec{q}\cdot \vec{\sigma}_2 ) \nonumber\\
&+& C_7 \, (\vec{k}\cdot \vec{\sigma}_1 )(\vec{k}\cdot \vec{\sigma}_2 )~.
\eea
Note that the overall UV regulator has been neglected as discussed in Section \ref{sec:HF}, and the $\Lambda$-dependence of the couplings is suppressed for brevity. For these contributions to the Hartree-Fock energy, there is no need to perform the DME since the zero-range nature of $V^{\text{NN}}_{ct}$ results in an expression that is already in the form of a bilinear Skyrme-like EDF with density-independent coupling constants
(see Appendix~\ref{appendix:contacts}). Consequently, each DME coupling at the HF level can be decomposed as the sum of a density-independent, $\Lambda$-dependent piece coming from $V^{\text{NN}}_{ct}(\Lambda)$ and a density-dependent, $\Lambda$-independent piece coming from finite-range pion exchanges, i.e.

\begin{equation}
C_t^{\rho\tau} \equiv C_t^{\rho\tau}(\Lambda;V^{\text{NN}}_{ct}) + C_t^{\rho\tau}(\Rvec;V^{\text{NN}}_{\pi})\,\,,\,\, \rm{etc.} \label{split}
\end{equation}

As discussed in the Introduction, while Hartree-Fock becomes a reasonable zeroth-order approximation with low-momentum interactions, it is necessary to go to at least 2nd-order in MBPT to obtain a reasonable description of bulk properties of infinite nuclear matter as well as binding energies and charge radii of closed-shell nuclei. However, a consistent extension of the DME beyond the Hartree-Fock level to treat the state-dependent energy denominators that arise has not been formulated to the best of our knowledge. At this point in time, any attempt to construct a {\it quantitative} Skyrme-like EDF starting from microscopic many-body theory must inevitably resort to unsystematic approximations (e.g., average state-independent energy denominators) in performing the DME on iterated contributions beyond the HF level~\cite{negele72,negele75,hofmann97,Kaiser:2002jz,Kaiser:2009me}, and/or to the re-introduction of some phenomenological parameters to be adjusted to data. In the present paper, we are motivated by the following observations
\begin{enumerate}
\item A more quantitative many-body treatment such as 2nd-order MBPT or the Brueckner-Hartree-Fock (BHF) approximation can be approximately cast into the same form as Eqs.~\ref{eq:finalH} and \ref{eq:finalF} if one neglects both the state-dependence of the intermediate-state energy denominators and the non-locality of the corresponding $G$ matrix vertex.
\item The in-medium $G$ matrix vertex differs from the free-space NN potential mostly at short distances and ``heals'' to the free-space potential at sufficiently large distances.  In the zeroth approximation, this amounts to a $k_F(\Rvec)$-dependent renormalization of the couplings for the contact interaction $V^{\text{NN}}_{ct}$.
\item The coupling constants $\{C_S, C_T, C_1,\ldots\}$ of $V^{\text{NN}}_{ct}$ are usually matched to low-energy NN scattering data and deuteron properties, although in principle they could be matched to low-energy finite nuclei data.
\end{enumerate}

Based on these observations, we advocate a semi-phenomenological approach in which the DME couplings from the finite-range Fock energy contributions are added to empirical Skyrme EDFs whose parameters are then re-fit to nuclear matter and finite nuclei properties. Although the treatment of the N$^2$LO NNN contribution to the HF energy is postponed to a separate paper, the DME couplings from its long-range part constitute an integral part to be added to phenomenological Skyrme parameters. Those re-fit parameters can thus be viewed as containing the effects of the HF contribution from the contact interactions $V^{\text{NN}}_{ct}$ plus higher order effects that would arise in a more sophisticated BHF or 2nd-order MBPT calculation. Finally, due to the loose connection of the refit Skyrme parameters to the EFT contact terms, the EFT concept of naturalness~\cite{Furnstahl:1997hq} might provide useful theoretical constraints for the fitting procedure. The first calculations following this semi-phenomenological approach are underway and will be the subject of a future work~\cite{stoitsov09a}.

\subsection{Master Formulas}
\label{subsec:master}
Inserting Eqs.~\ref{eq:DMEscalar} and~\ref{eq:DMEvector} into the Fock energy (Eq.~\ref{eq:finalF}) and performing tedious but straightforward algebra to cast the expression into the same form as Eq.~\ref{eq:EDF}, we  obtain the following ``master formulas'' for the DME couplings:

\bea
\label{eq:master1}
C_t^{\rho\rho}&=& -\frac{1}{\pi k_F^3}\,\int\! q^2dq \,\Gamma_c^{xt}(q)\,\bigl[\,I_1(q/k_F) \,+\, \nonumber\\
&&\hspace{1.4in}+\,\, \frac{1}{5}I_2(q/k_F)\bigr] \\
C_t^{\rho\tau}&=& \frac{1}{3\pi k_F^5}\,\int\! q^2dq\, \Gamma_c^{xt}(q)\,I_2(q/k_F) \\
C_t^{\rho\Delta\rho} &=& -\frac{1}{4}C_t^{\rho\tau} \\
C_t^{JJ} &=& - \frac{1}{4\pi k_F^3}\,\int q^2dq\, I_3(q/k_F)\,\bigl( 1 + \frac{2}{3}q\partial_q \bigr)\,\Gamma_T^{xt}(q)\nonumber\\
                     &&\hspace{.3in}-\,\,\frac{1}{6\pi k_F^5}\int q^2dq\, \Gamma_S^{xt}(q)\,I_4(q/k_F) \,,
\label{eq:master2}
\eea
where terms with more than two gradients have been dropped, although investigating the effect of higher-order gradients might be of interest in the future~\cite{Gebremariam:2009ff}. The $I_n(q/k_F)$ functions are defined as
\bea
I_1(\bar{q}) &\equiv& \int x^2 dx\, j_0(\bar{q}x)\,\bigl(\Pi^{\rho}_0(x)\bigr)^2 \, \\
I_2(\bar{q}) &\equiv& \int x^4 dx\, j_0(\bar{q}x)\,\Pi^{\rho}_0(x)\,\Pi^{\rho}_2(x) \, \\
I_3(\bar{q}) &\equiv& \int x^2 dx\, j_0(\bar{q}x)\,\bigl(\Pi^{\svec}_1(x)\bigr)^2 \, \\
I_4(\bar{q}) &\equiv& \int x^4 dx\, j_0(\bar{q}x)\,\bigl(\Pi^{\svec}_1(x)\bigr)^2 \,.
\eea
Inserting the NV-DME and PSA-DME expressions for the $\Pi$-functions gives
\bea
I_1(\bar{q}) \!&=&\!\frac{3\,\pi}{32}\, \bigl(\bar{q}^3 \,-\, 12\bar{q} \,+\,16\bigr)\,\theta(2-\bar{q}) \,  \\
I_2(\bar{q}) \!&=&\!  -\frac{35\,\pi}{128}\bigl(\bar{q}^5-18\bar{q}^3+40\bar{q}^2-24\bar{q}\bigr)\theta(2-\bar{q}) \, \\
I_3(\bar{q})\!&=&\! \frac{\pi}{4q}\,\theta(2-\bar{q}) \\
I_4(\bar{q})\!&=&\! -\frac{\pi}{4q}\,\Bigl( \frac{\partial}{\partial\bar{q}}\delta(\bar{q}) -\frac{\partial}{\partial\bar{q}}\delta(\bar{2-q})\Bigr)
\eea
for the NV-DME and
\bea
I_1(\bar{q})\! &=&\! I_3(\bar{q}) = \frac{3\,\pi}{32}\,\bigl(\bar{q}^3 \,-\, 12\bar{q} \,+\,16\bigr)\,\theta(2-\bar{q}) \,  \\
I_2(\bar{q}) \!&=&\! I_4(\bar{q}) =  \frac{9\pi}{8\bar{q}}\,\bigl(2 - \bar{q}^2\bigr)\,\theta(2-\bar{q})\,
\eea
for the PSA-DME, respectively.
\subsection{Skeleton Expressions}

The lengthy analytic expressions for the DME couplings obtained from the master formulas tend to obscure their underlying structural simplicity. Therefore, it is more illuminating to display the couplings in ``skeleton form'' and relegate the explicit expressions to the Mathematica notebook and Appendix~\ref{appendix:couplings}. Each coupling $C^{g}_t$ is given by the sum of the LO ($n=0$), NLO ($n=1$), and N$^2$LO ($n=2$) contributions
\begin{equation}
C^{g}_t(u)\, =\, \sum_{n= 0}^{2}C^{g}_{t,n}(u)\quad g \in \{\rho\rho, \rho\tau, \rho\Delta\rho,\ldots\}\,,
\end{equation}
with $u\equiv k_F/m_{\pi}$ and
\begin{equation}
C^{g}_{t,n}(u) = \sum_{j=0}^2 \alpha^{g}_j(n,t,u)\mathcal{F}_j(n,u)\,\, ,
\end{equation}
where $\alpha^{g}_j(n,t,u)$ are rational polynomials in $u$ and $\mathcal{F}_j(n,u)$ are functions which may exhibit non-analytic behavior  in $u$ due to the finite-range of the NN interaction.

We note that the detailed form of the skeleton NLO and N$^2$LO expressions depends on the value of the spectral function regulator (SFR) mass $M_{{\rm sfr}\,\,}$ used to regulate the divergent loop integrals of the two-pion exchange potentials (TPEP), see Appendix~\ref{appendix:NNpot} and Ref.~\cite{Epelbaum:2004fk}. For simplicity, the following skeleton expressions were obtained for $M_{{\rm sfr}\,\,}\rightarrow\infty$, which corresponds to using dimensional regularization for the loop integrals that enter into the expressions for the NLO and N$^2$LO two-pion exchange potentials as in Ref.~\cite{Entem:2003ft}. In the skeleton expressions listed below, we use a more compact notation where the dependence on $u$, $t$, and $n$ is not explicitly shown for the $\alpha$'s:

\begin{flushleft}
$\bullet\quad$\underline{\em LO}\\

\bea
C^{g}\!&=&\!\alpha^{g}_0 + \alpha^{g}_1\log\bigl(1+4u^2\bigr)
+\alpha^{g}_2\arctan(2u)
\eea
\end{flushleft}

\begin{flushleft}
$\bullet\quad$\underline{\em NLO}\\

\bea
C^{g}& = &\alpha^{g}_0 + \alpha^{g}_1\biggl[\log\bigl(1+2u^2+2u\sqrt{1+u^2}\bigr) \biggr]^2\nonumber \\
&&\hspace{-.2in}+\, \alpha^{g}_2\sqrt{1+u^2}\log\bigl(1+2u^2+2u\sqrt{1+u^2}\bigr)
\eea
\end{flushleft}

\begin{flushleft}
$\bullet\quad$\underline{\em N$^2$LO}\\
\bea
C^{g}\!&=&\!\alpha^{g}_0 + \alpha^{g}_1\log\bigl(1+u^2\bigr)
+\alpha^{g}_2\arctan(u)
\eea
\end{flushleft}
Note that rather similar results are obtained for both the PSA-DME and NV-DME, as well as for finite values of the SFR mass $M_{{\rm sfr}\,\,}$ in the TPEP. In the following section, we present results for the PSA-DME with $M_{{\rm sfr}\,\,}= 500$ MeV unless otherwise specified.

\section{Selected results}
\label{sec:Results}

\subsection{Density-dependent Fock couplings}
\label{fockdepdens}

In the present section, the non-trivial (isoscalar-) density dependence of Fock DME couplings $C_t^{g}(\Rvec;V^{\text{NN}}_{\pi})$, with $g \in \{\rho\rho, \rho\tau, \rho\Delta\rho,\nabla\rho\nabla\rho,J\nabla\rho,JJ\}$ is briefly analyzed. First, we note that $C_t^{J\nabla\rho}(\Rvec;V^{\text{NN}}_{\pi})=0$ through N$^2$LO since the two-body spin-orbit interaction is entirely carried by contact terms (see Eq.~\ref{Vcon}). Second, since we are restricting the DME described in Sec.~\ref{sec:DME} to the Fock energy contribution, one finds $C_t^{\nabla\rho\nabla\rho}(\Rvec;V^{\text{NN}}_{\pi})=0$.  Of course, the $\rho\Delta\rho$ terms can be transformed by partial integration into $(\nabla\rho)^2$ terms, but we chose not to do so since the $k_F(\Rvec)$-dependence of the couplings results in more complicated expressions. Conversely, we note that it is not possible to convert $(\nabla\rho)^2$ terms entirely to the $\rho\Delta\rho$ form when the couplings depend on $k_F(\Rvec)$ in contrast to usual Skyrme functionals.

\begin{figure}[t]
  \includegraphics[width=3.1in,clip=]{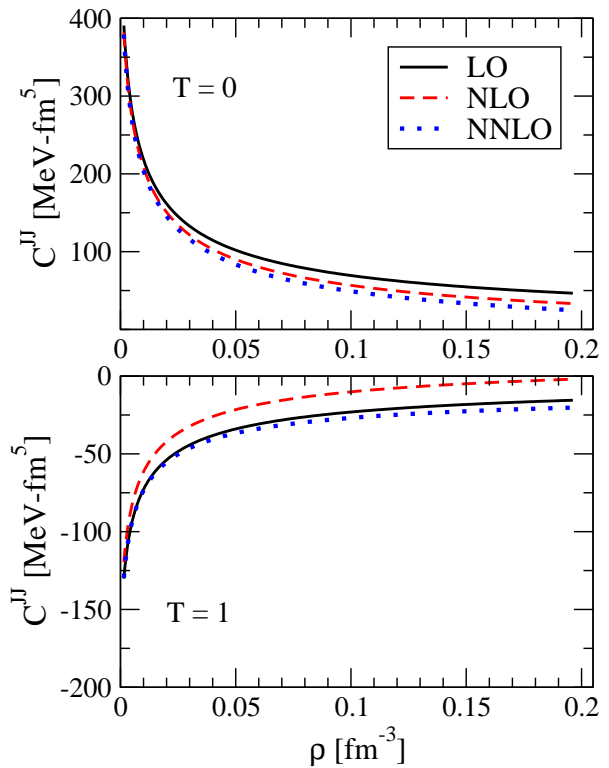}
\caption{Density dependence of $C_t^{JJ}(\Rvec;V^{\text{NN}}_{\pi})$ calculated through LO, NLO, and N$^2$LO for the isoscalar (upper panel) and isovector (lower panel) couplings. A SFR mass of $M_{{\rm sfr}\,\,}= 500$ MeV was used in the NLO and NNLO two-pion exchange potentials.}%
\label{fig:LONLON2LOJJ}
\end{figure}

\begin{figure}[t]
\includegraphics*[width=3.1in,clip=]{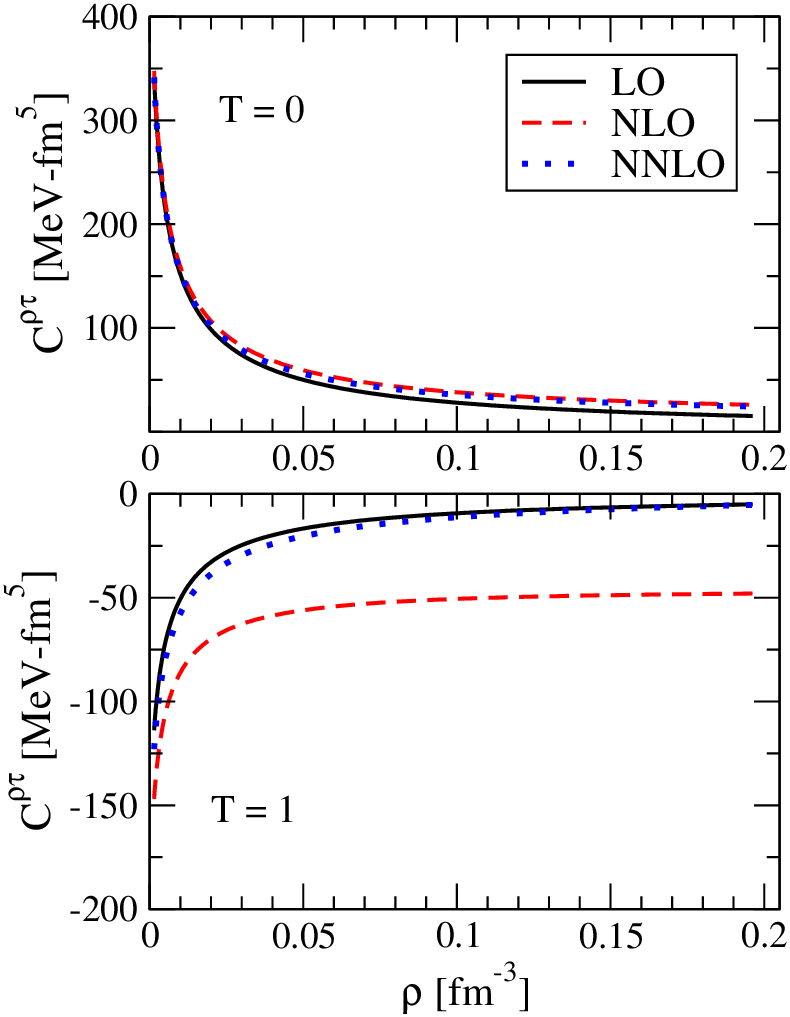}
\caption{Same as Fig.~\ref{fig:LONLON2LOJJ} for $C_t^{\rho\tau}(\Rvec;V^{\text{NN}}_{\pi})$. }%
\label{fig:LONLON2LO_rhotau}
\end{figure}

\begin{figure}
\includegraphics*[width=3.1in,clip=]{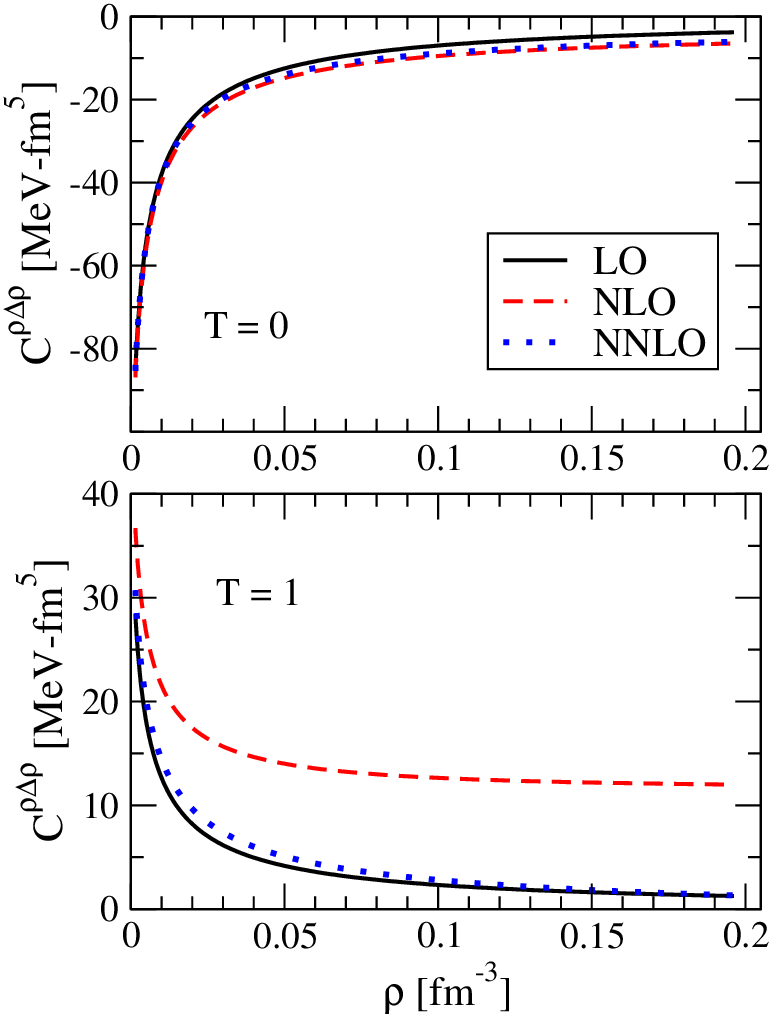}
\caption{Same as Fig.~\ref{fig:LONLON2LOJJ} for $C_t^{\rho\Delta\rho}(\Rvec;V^{\text{NN}}_{\pi})$. }%
\label{fig:LONLON2LO_rhodeltarho}
\end{figure}

\begin{figure}
\includegraphics*[width=3.1in,clip=]{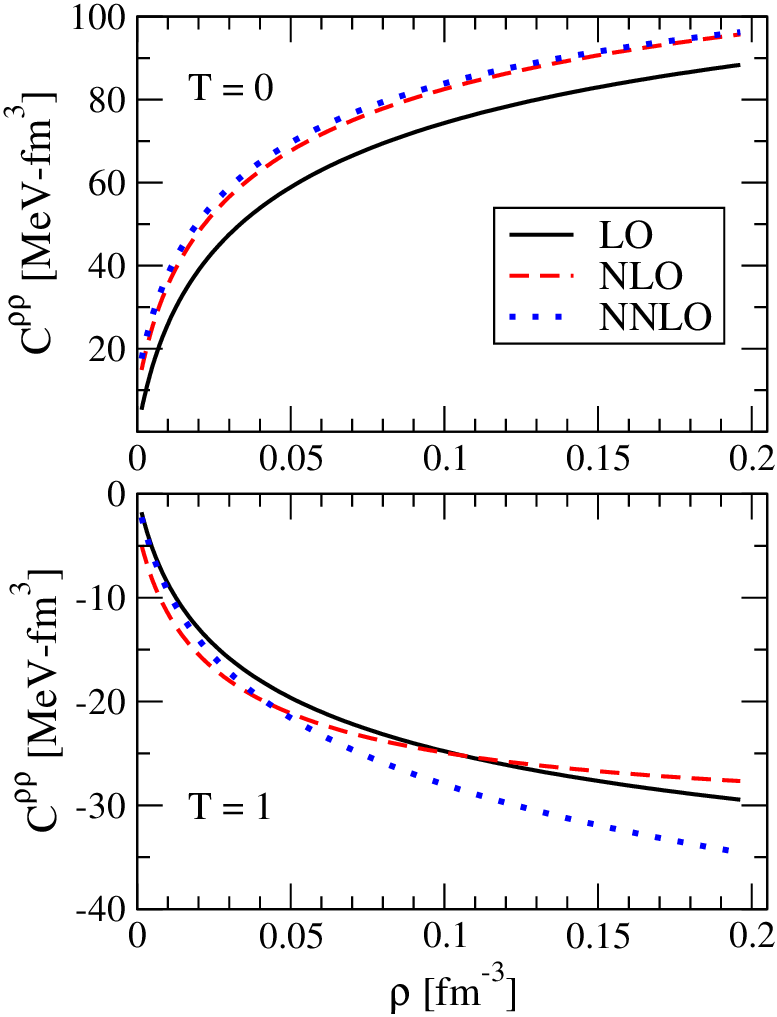}
\caption{Same as Fig.~\ref{fig:LONLON2LOJJ} for $C_t^{\rho\rho}(\Rvec;V^{\text{NN}}_{\pi})$.  }%
\label{fig:LONLON2LO_rhorho}
\end{figure}

Referring to Figs.~\ref{fig:LONLON2LOJJ}-\ref{fig:LONLON2LO_rhorho}, the non-zero isoscalar and isovector couplings are shown including LO, NLO and N$^2$LO contributions. The main feature to extract from these results is that the non-trivial density-dependence is controlled by the longest-range parts of the NN interaction. Indeed, the density profile of the couplings is driven almost entirely by the LO term (one-pion exchange) whereas the NLO and N$^2$LO interactions that are built from shorter-range two-pion exchanges provide small corrections to the LO density-dependence. One might be surprised by the fact that the successive contributions do not exactly follow the hierarchy LO $>$ NLO $ >$ N$^2$LO. However, this is somewhat misleading since only the Fock contributions from
finite-range NN pieces are actually shown and cannot be expected to maintain such a hierarchy without including
N$^2$LO NNN and NN-contact contributions.

The second important result is that all non-zero couplings exhibit a substantial isoscalar density-dependence\footnote{Note the large y-axis scale used in most of the figures. The reason that the vertical scales are large becomes clear as one refers to typical values of the couplings for standard Skyrme parameterizations; see section~\ref{compskyrme}.}, especially as one goes to small densities.  Such in-medium dependencies are obviously at variance with standard phenomenological Skyrme parameterizations for which only $C_t^{\rho\rho}$ depends on the medium in time-reversal invariant systems. Investigating the impact of such non-trivial in-medium dependencies generated by pion exchanges is one of the primary long-term goals of our project. Of course, many-body correlations generated at higher-orders from both short-range-contact and long-range-pion physics will provide couplings with additional medium dependencies. As discussed in the next section, the impact of such in-medium effects on observables and Skyrme phenomenology can only be fully characterized by performing full-fledged EDF calculations.

\subsection{Comparison with Skyrme phenomenology}
\label{compskyrme}

According to Eq.~\ref{split}, Fock DME couplings $C_t^{g}(\Rvec;V^{\text{NN}}_{\pi})$ must be complemented with the contribution $C_t^{g}(\Lambda;V^{\text{NN}}_{ct})$ whose expressions are given in Appendix~\ref{appendix:contacts}. As already discussed, three additional types of contributions to the EDF should be further considered. First is the Hartree term given by Eq.~\ref{eq:finalH}. Although such a contribution solely depends on the local part of the density matrix, it does not take the form of a local Skyrme-like EDF (Eq.~\ref{eq:EDF}) when treated exactly, as advocated here. Second are the Hartree-Fock contributions from both the long-range ($V^{\text{3N}}_{\pi}$) and short-range ($V^{\text{3N}}_{ct}$) parts of the chiral NNN at N$^2$LO. The application of the DME to the $V^{\text{3N}}_{\pi}$ contributions, which eventually leads to additional density-dependent contributions $C_t^{g}(\Rvec;V^{\text{3N}}_{\pi})$, will be discussed in a separate paper. Last but not least, contributions beyond HF, which can hopefully be recast into a quasi-local EDF form, will add an additional in-medium renormalization to the couplings.

Keeping in mind the above warnings as to what a complete non-empirical EDF should contain, we now perform a primitive comparison with Skyrme phenomenology with the goal of providing a zeroth-order assessment of the non-trivial in-medium dependence of the coupling functions. To do so, we compensate for all missing pieces beyond Fock DME couplings $C_t^{g}(\Rvec;V^{\text{NN}}_{\pi})$ by providing an "uncertainty band" generated by imposing naturalness requirements for the coupling {\it constants} of the associated bilinear term in the EDF.  In the current context, naturalness means that the (dimensionless) coupling constants are of order unity after appropriate combinations of the strong interaction scales $f_{\pi}$ and $\Lambda_{\chi}$ are extracted from the energy density\footnote{We use $f_{\pi}=93$ MeV and $\Lambda_{\chi}=700$ MeV when generating the NDA estimates.}. We refer to Ref.~\cite{Furnstahl:1997hq} for details on applying the naive dimensional analysis (NDA) of Manohar and Georgi~\cite{Manohar:1983md}  to Skyrme EDFs. The resulting ``naturalness band'' should only be viewed as an order-of-magnitude estimate of the missing higher-order pieces, which in any event will carry additional non-trivial density dependencies in a fully microscopic approach. Still, the fact that most phenomenological Skyrme parameterizations conform to these naturalness bounds ~\cite{Furnstahl:1997hq} provides some justification for such a primitive procedure.

\begin{figure}
  \includegraphics*[width=3.1in,clip=]{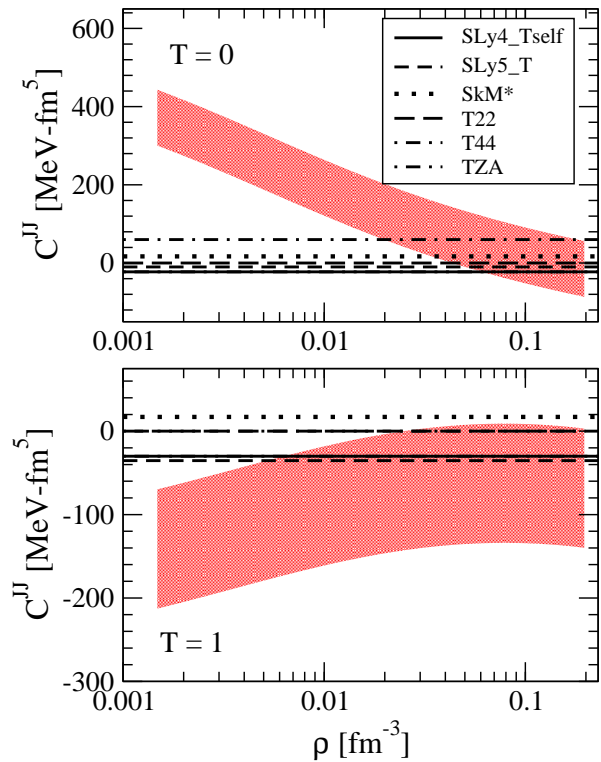}
\caption{Fock DME coupling $C_t^{JJ}(\Rvec;V^{\text{NN}}_{\pi})$ augmented with a "natural" Skyrme-like contribution (see text) and compared to the corresponding coupling from a representative set of Skyrme parameterizations. Upper (lower) panel: isoscalar (isovector) coupling.}%
\label{bandJJ}
\end{figure}

\begin{figure}
\includegraphics*[width=3.1in,clip=]{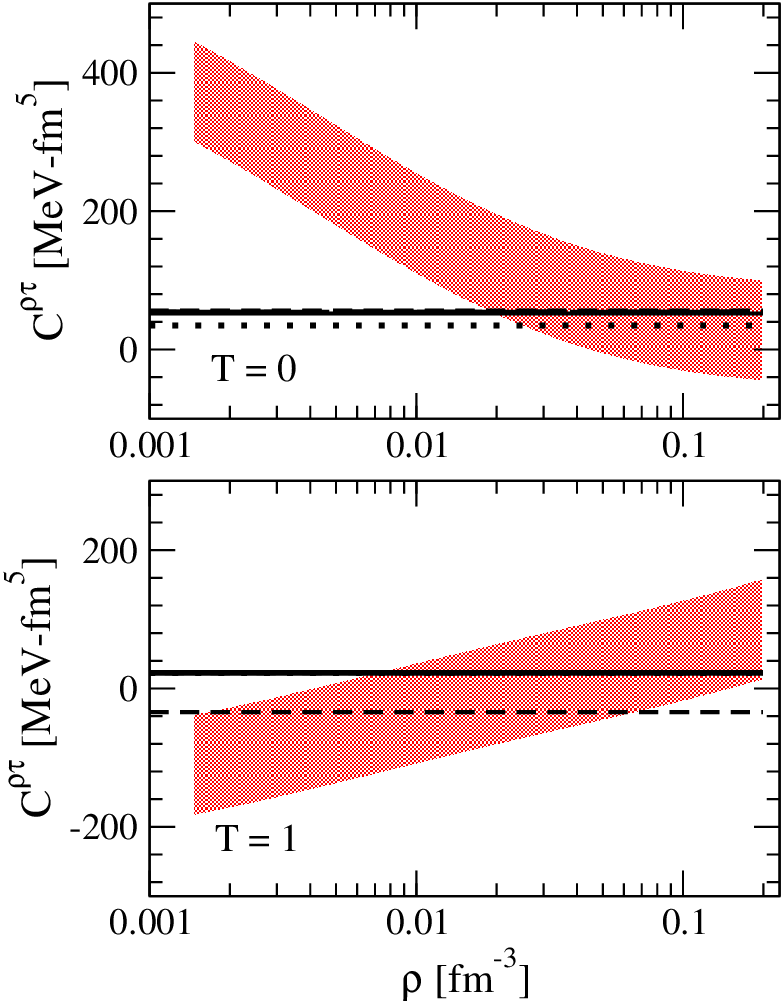}
\caption{Same as Fig.~\ref{bandJJ} for $C_t^{\rho\tau}(\Rvec;V^{\text{NN}}_{\pi})$.}%
\label{bandrhotau}
\end{figure}

\begin{figure}
\includegraphics*[width=3.1in,clip=]{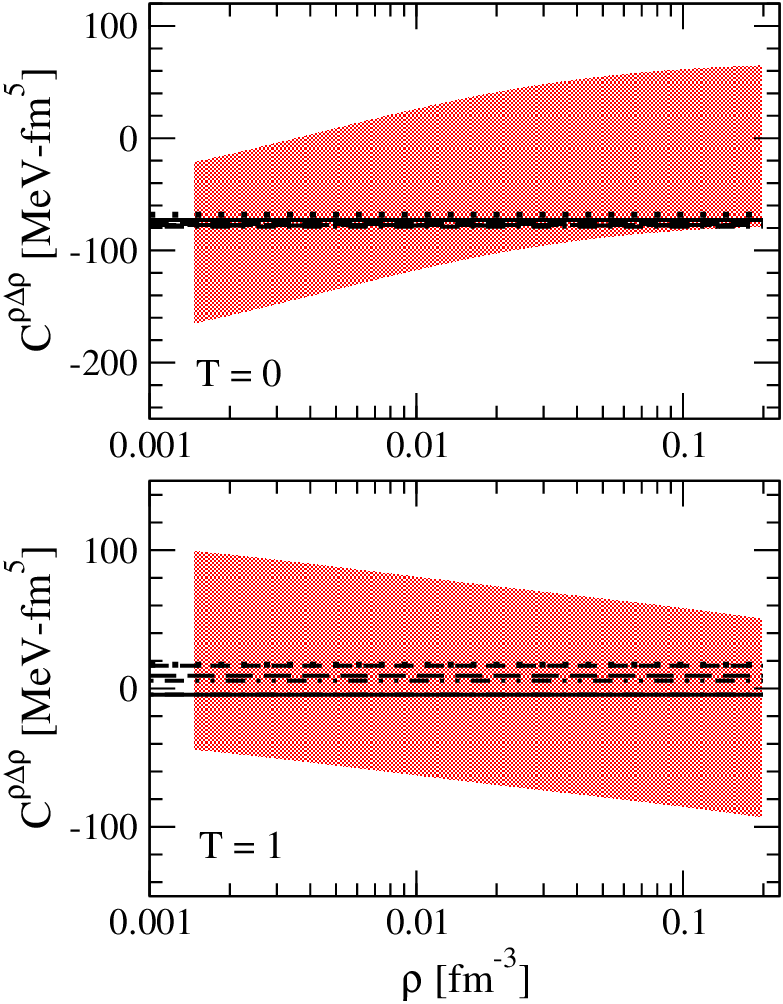}
\caption{Same as Fig.~\ref{bandJJ} for $C_t^{\rho\Delta\rho}(\Rvec;V^{\text{NN}}_{\pi})$.}%
\label{bandrhodeltarho}
\end{figure}

\begin{figure}
\includegraphics*[width=3.1in,clip=]{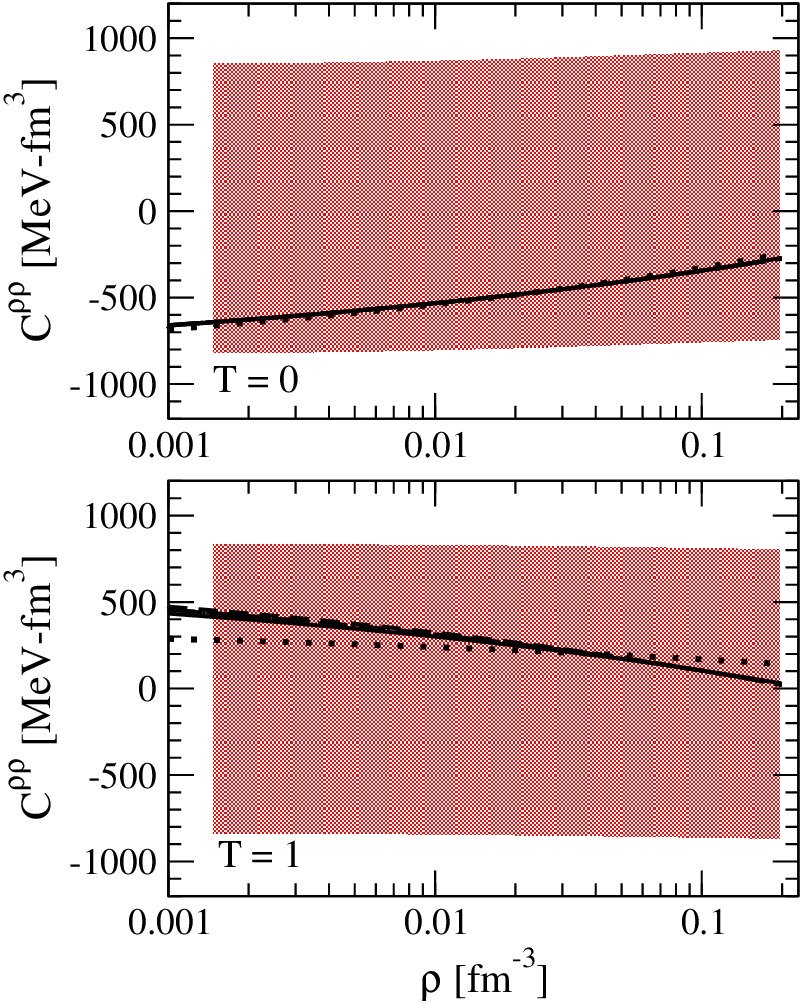}
\caption{Same as Fig.~\ref{bandJJ} for $C_t^{\rho\rho}(\Rvec;V^{\text{NN}}_{\pi})$.}%
\label{bandrhorho}
\end{figure}

Given that $C_t^{J\nabla\rho}(\Rvec;V^{\text{NN}}_{\pi})=0$ through N$^2$LO,  we do not provide a graphical comparison to Skyrme phenomenology since such couplings are entirely provided by the ``natural'' contribution (i.e., the NN spin-orbit interaction is a contact interaction). Note, however, that the NNN force at N$^2$LO has long-range spin-orbit contributions that will provide non-trivial (i.e. density-dependent) spin-orbit couplings in the EDF.  All other isoscalar and isovector couplings are shown in Figs.~\ref{bandJJ}-\ref{bandrhorho}. Compared to the previous section, a logarithmic scale is used for the horizontal axis that gives visually more weight to lower densities.

To perform the comparison, we employ a representative set of modern Skyrme parameterizations: SkM$^\ast$~\cite{bartel82a}, T22~\cite{Lesinski:2007zz}, T44~\cite{Lesinski:2007zz}, TZA~\cite{Bender:2009ty}, SLy4T$_\text{self}$~\cite{Bender:2009ty}, SLy5+T~\cite{colo07a}. Except for SkM$^\ast$, they all result from recent investigations that aimed at pinning down (while keeping the overall quality of modern parameterizations) values of the tensor couplings $C_t^{JJ}$.  The Skyrme couplings displayed in Figs.~\ref{bandJJ}-\ref{bandrhorho} are essentially all within the uncertainty band of the DME-inspired coupling functions around saturation density. Although only qualitative, this is a very significant, i.e. non-obvious, result. Of course, the rather conservative uncertainty-band used is typically larger than the difference between various fine-tuned parameterizations. A more systematic non-empirical treatment of all contributions to the coupling functions is expected to narrow down the uncertainty band and perhaps allow one to discriminate between various empirical parameterizations. Of particular interest are tensor couplings $C_t^{JJ}$ whose preferred range\footnote{The definition of $C_t^{JJ}$ presently used corresponds to half the one of Ref.~\cite{Lesinski:2007zz}.} is not settled yet by the phenomenology (see Fig.~\ref{bandJJ}).

As already stressed, the present status of our approach is such that the above comparison should be taken as qualitative at best. Only minimal quantitative information can be gleaned from comparisons at the level of the EDF couplings as (i) only the total energy is observable and that (ii) self-consistent effects can be significant, giving rather different final results even if one starts with parameterizations whose couplings look alike on the scale of Figs.~\ref{bandJJ}-\ref{bandrhorho}. Eventually, comparisons based on fully self-consistent EDF calculations performed with the (semi-)non-empirical energy functionals presented here will provide useful quantitative information. Such a project is currently underway~\cite{stoitsov09a}.

\begin{figure}
  \includegraphics[width=3.1in,clip=]{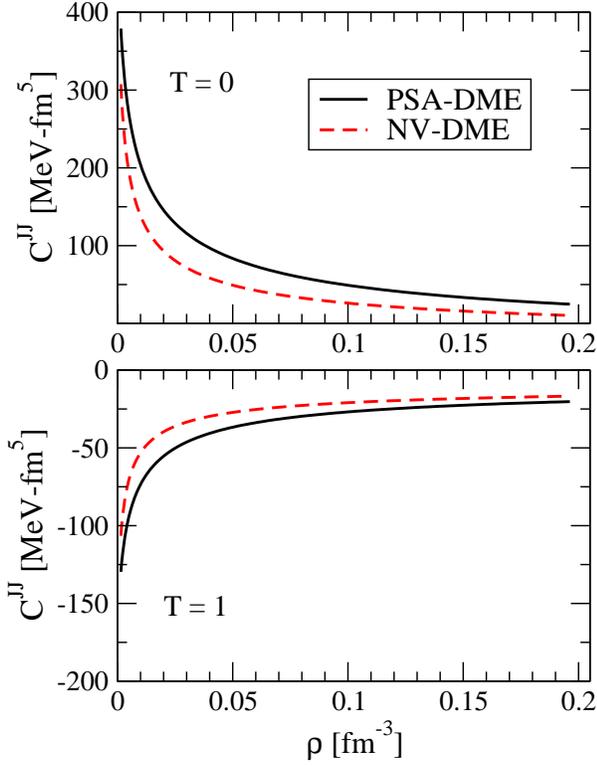}
\caption{Density dependence of the PSA-DME and NV-DME couplings $C_t^{JJ}(\Rvec;V^{\text{NN}}_{\pi})$ calculated through N$^2$LO for the isoscalar (upper panel) and isovector (lower panel) couplings}%
\label{fig:PSANVJJ}
\end{figure}

\begin{figure}
\includegraphics*[width=3.1in,clip=]{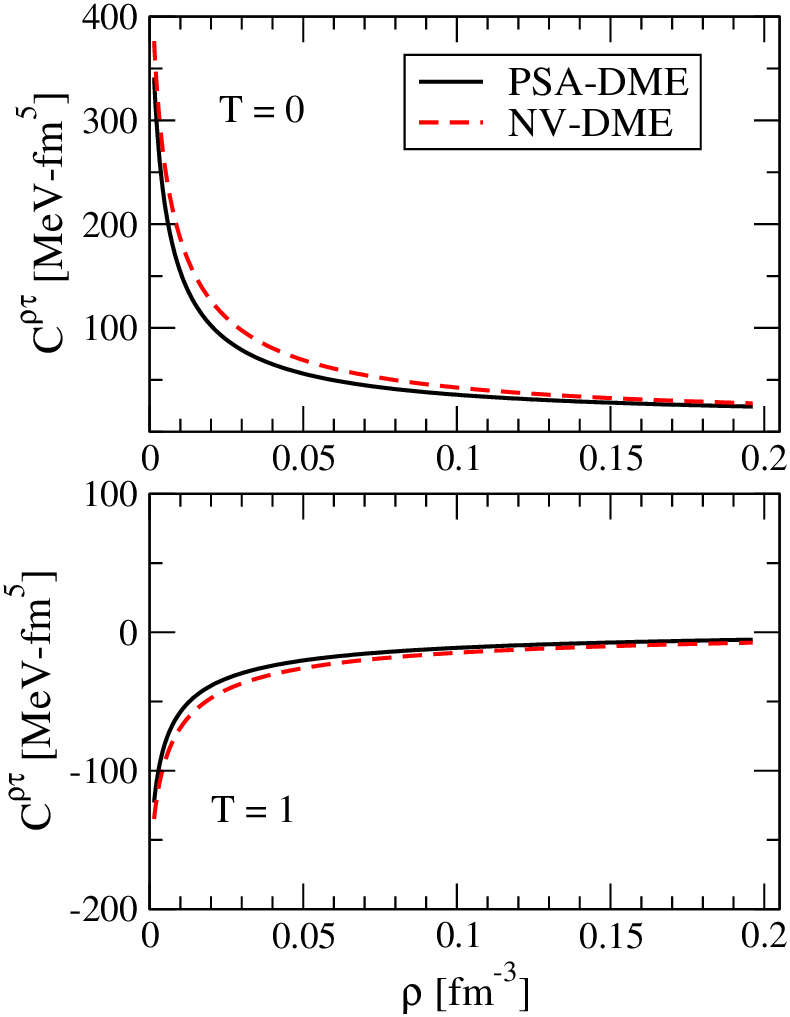}
\caption{Same as Fig.~\ref{fig:PSANVJJ} for $C_t^{\rho\tau}(\Rvec;V^{\text{NN}}_{\pi})$. }%
\label{fig:PSANVrhotau}
\end{figure}

\begin{figure}
\includegraphics*[width=3.1in,clip=]{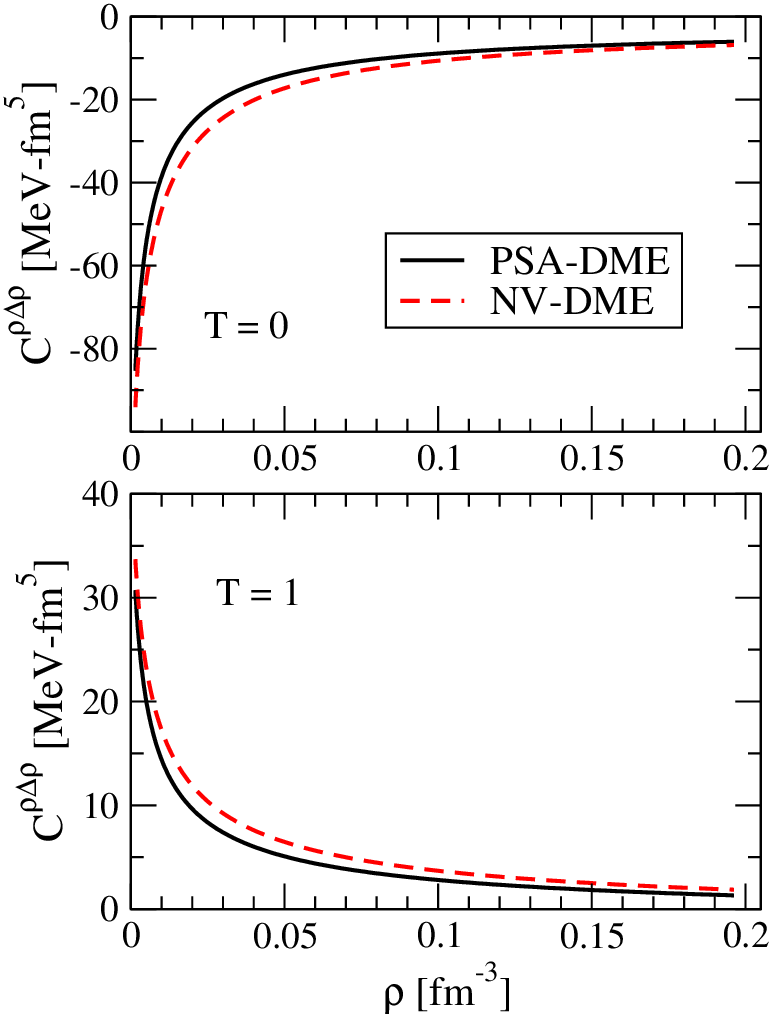}
\caption{Same as Fig.~\ref{fig:PSANVJJ} for $C_t^{\rho\Delta\rho}(\Rvec;V^{\text{NN}}_{\pi})$. }%
\label{fig:PSANVrhodeltarho}
\end{figure}

\begin{figure}
\includegraphics*[width=3.1in,clip=]{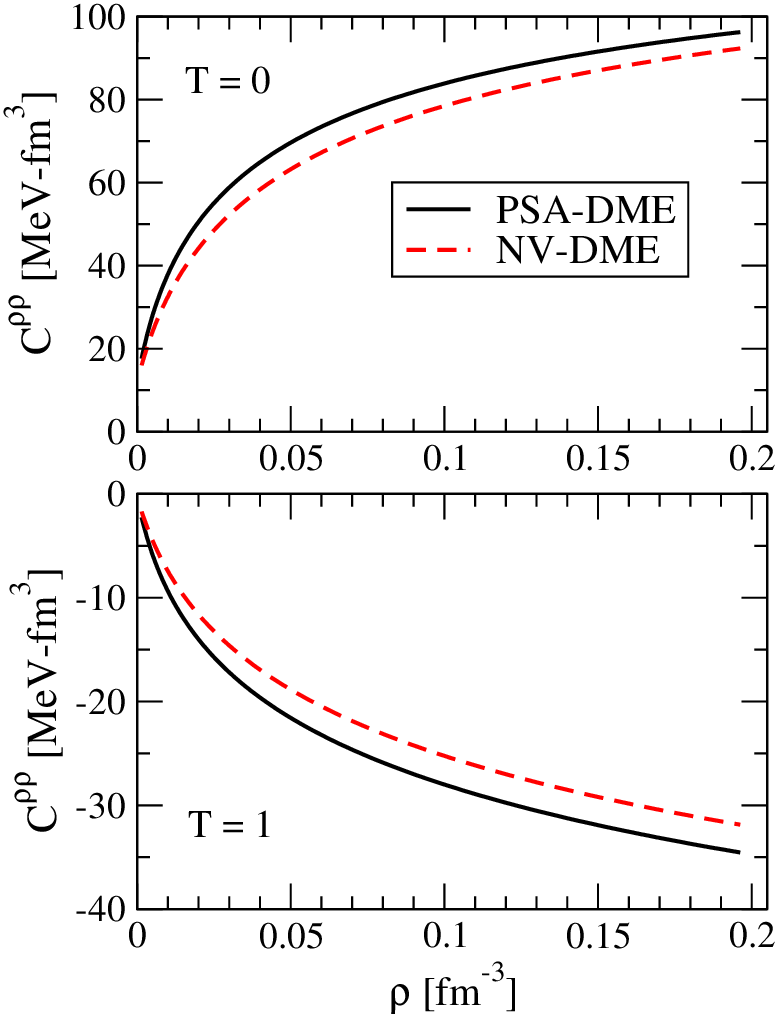}
\caption{Same as Fig.~\ref{fig:PSANVJJ} for $C_t^{\rho\rho}(\Rvec;V^{\text{NN}}_{\pi})$.  }%
\label{fig:PSANVrhorho}
\end{figure}

Lastly, Figs.~\ref{fig:PSANVJJ}-\ref{fig:PSANVrhorho} compare the various couplings obtained using the original NV-DME and the PSA-DME used here. It is reassuring that, while the numerical values of the couplings can change somewhat depending on which variant of the DME is used, the overall density profiles are relatively insensitive to this choice. At least within the semi-phenomenological approach outlined in the present paper, the similar density profiles for the PSA-DME and NV-DME couplings implies that full-fledged EDF calculations should be fairly insensitive to this choice since refitting the Skyrme constants largely absorbs these (approximately density-independent) differences.
\section{Summary and Conclusions}
\label{sec:Conclusions}

It is by now well-established that empirical Skyrme functionals (in present form) exhibit critical limitations that are often manifested by parametrization-dependent predictions away from known data, e.g., see Ref.~\cite{Duguet:2006jg,lesinski06a}. One possible remedy is to use many-body perturbation theory and knowledge of the underlying two- and three-nucleon interactions to identify missing microscopic long-distance physics, together with density matrix expansion techniques to incorporate these missing ingredients into existing Skyrme machinery~\cite{Lesinski:2008cd,Drut:2009ce,Duguet:2009gc,Bogner:2008kj}.

In the present work, our primary focus is approximating the spatial non-locality of the Fock contribution from the chiral N$^2$LO NN interaction, while keeping the Hartree contribution exact and postponing the treatment of the NNN part to a separate paper. The generic structure of the chiral interactions is such that each DME coupling $C_t^{g}$ ($t=0,1$ and  $g \in \{\rho\rho, \rho\tau, \rho\Delta\rho,J\nabla\rho,JJ\}$) decomposes into a cutoff-dependent coupling constant $C_t^{g}(\Lambda;V^{\text{NN}}_{ct})$ arising from the zero-range contact interactions and a cutoff-independent coupling {\it function} $C_t^{g}(\Rvec;V^{\text{NN}}_{\pi})$ of the density arising from the long-range pion exchanges. In the short term, this separation (which is completely unambiguous at the Hartree-Fock level) between long- and short-distance physics motivates adding the $C_t^{g}(\Rvec;V^{\text{NN}}_{\pi})$ to existing Skyrme functionals, upon which the original Skyrme constants can then be refit to data. This semi-phenomenological approach is motivated by the observation that the EFT contact terms can in principle be fixed to any low-energy quantities. Therefore, refitting the Skyrme constants can be viewed as matching a microscopically-constrained Skyrme-like functional containing explicit pion physics (albeit in a zeroth-order HF approximation at this point) to finite-density observables.

Restricting ourselves to time-reversal invariant systems, analytical expressions of the Fock DME couplings $C_t^{g}(\Rvec;V^{\text{NN}}_{\pi})$ have been derived as a function of the isoscalar density. The rather lengthy analytic expressions can be downloaded through a companion Mathematica notebook, and are also collected in Appendix~\ref{appendix:couplings}. The novel density-dependencies of the couplings are controlled by the longest-ranged parts of the NN interaction, which implies that these microscopic constraints are coming from the best-understood parts of the underlying nuclear interactions. The dependence on the isoscalar density is significant for all couplings over the density interval of interest, which is obviously at variance with standard phenomenological Skyrme parameterizations whose only density-dependent couplings are $C_t^{\rho\rho}$ $(t=0,1)$. In the longer term, investigating the impact of such non-trivial in-medium dependencies generated by pion exchanges is one of the central goals of the present project~\cite{stoitsov09a}.

The rich spin and isospin dependence of the chiral EFT one- and two-pion-exchange interactions gives us hope that their inclusion will provide valuable microscopic constraints on the poorly-understood isovector properties of the EDF.  We do not expect dramatic changes for bulk nuclear properties due to the tendency of pions to average out in spin and isospin sums, but we do expect interesting consequences for single-particle properties (which phenomenology tells us are sensitive probes of the tensor force) and systematics along long isotope chains (which should be sensitive to the isovector physics coming from pion-exchange interactions).  Another potentially significant advantage of the current approach is that two very different microscopic origins of spin-orbit properties (i.e., short-range NN and long-range NNN spin-orbit interactions) are treated on equal footing. This is in contrast to empirical Skyrme and Gogny functionals, where the zero-range spin-orbit interaction has no obvious connection with the sub-leading (but quantitatively significant) NNN sources of spin-orbit splittings. In a forthcoming paper, we will extend the DME to include HF contributions from chiral three-nucleon interactions at N$^2$LO. Such an extension will allow one to probe the impact of microscopic three-nucleon forces on the structure of medium- and heavy-mass nuclei.

The EDF obtained as a result of the present paper and the subsequent NNN paper only contains non-empirical Hartree-Fock contributions, such that further correlations must be added to produce any reasonable description of nuclei. As explained above, in the short term this will be done empirically by adding the DME couplings to empirical and Skyrme functional whose coupling constants are refitted to data~\cite{stoitsov09a}. While this is an admittedly empirical procedure, it is motivated by the well-known observation that a Brueckner G-matrix differs from the vacuum NN interaction only at short distances. Therefore, one can interpret the refit to data as approximating the short-distance part of the G-matrix with a zero-range expansion through second order in gradients. Eventually though, it is the goal of a future work to design a generalized DME that is suited to higher orders in perturbation theory~\cite{rotival09a}.

\begin{acknowledgments}
We thank Dick Furnstahl and Jacek Dobaczewski for useful discussions.
This work was supported in part by the U.S. Department of Energy UNEDF SciDAC Collaboration under Contract No. DEFC02-07ER41457, and by the U.S. National Science Foundation under Grant No. PHY-0758125.
\end{acknowledgments}

\appendix

\section{Chiral EFT NN potentials}
\label{appendix:NNpot}
The NN potential in chiral EFT can be schematically written as
\bea
V^{\text{NN}} &=& V^{\text{NN}}_{1\pi} + V^{\text{NN}}_{2\pi} +\ldots+ V^{\text{NN}}_{ct}(\Lambda)\,,
\,\eea
where $V_{1\pi}$ and $V_{2\pi}$ are finite-range one- and two-pion exchange interactions dictated by the spontaneously broken chiral symmetry of QCD, and $V^{\text{NN}}_{ct}$ denotes the scale-dependent contact terms that encode the effects of integrated out degrees of freedom (heavier meson exchanges, high-momentum two-nucleon states, etc.) on low energy physics. As discussed in the text, our primary focus is on the finite-range pion-exchange interactions since they drive the non-trivial density-dependencies introduced by the DME. In the notation of Eq.~\ref{eq:EFTdecomp}, we list the non-zero finite-range contributions thru N$^2$LO:

\begin{widetext}
\begin{flushleft}
$\bullet\quad$\underline{\em LO}\\
\end{flushleft}
\beq
W_T(q) = -\Bigl(\frac{g_A}{2f_{\pi}}\Bigr)^2\, \frac{1}{q^2 + m_{\pi}^2}
\eeq

\begin{flushleft}
$\bullet\quad$\underline{\em NLO}\\
\end{flushleft}
\bea
W_C(q) &=& -\frac{1}{384\pi^2f_{\pi}^4}\,L^{M_{{\rm sfr}\,\,}}(q)\,\Biggl\{ 4m_{\pi}^2(5g_A^4 - 4g_A^2 -1)\,+\,q^2(23g_A^4 - 10g_A^2 -1) + \frac{48g_A^4m{\pi}^4}{4m_{\pi}^2 + q^2}\Biggr\}\\
V_T(q) &=& -\frac{1}{q^2}V_S(q) =  -\frac{3g_A^4}{64\pi^2f_{\pi}^4}L^{M_{{\rm sfr}\,\,}}(q)\\
\nonumber
\eea

\begin{flushleft}
$\bullet\quad$\underline{\em N$^2$LO}\\
\end{flushleft}
\bea
V_C(q) &=& -\frac{3g_A^2}{16\pi f_{\pi}^4}\Biggl\{2m_{\pi}^2(2c_1 - c_3) - c_3\,q^2\Biggr\}\,(2m_{\pi}^2 + q^2)\,A^{M_{{\rm sfr}\,\,}}(q) \\
W_T(q) &=& -\frac{1}{q^2}W_S(q) = -\frac{g_A^2}{32\pi f_{\pi}^4}\,c_4\,(4m_{\pi}^2 + q^2)\,A^{M_{{\rm sfr}\,\,}}(q)\,,\\
\nonumber
\eea
where the NLO and N$^2$LO loop functions are given by

\bea
L^{M_{{\rm sfr}\,\,}}(q) &=& \theta(M_{{\rm sfr}\,\,}-2m_{\pi})\,\frac{\omega}{2q}\ln{\frac{(M_{{\rm sfr}\,\,}\omega + qs)^2}{4m_{\pi}^2(M_{{\rm sfr}\,\,}^2 + q^2)}\,,\quad \omega = \sqrt{q^2+4m_{\pi}^2}, \quad s = \sqrt{M_{{\rm sfr}\,\,}^2 - 4m_{\pi}^2}\,}\,,\\
A^{M_{{\rm sfr}\,\,}}(q) &=& \theta(M_{{\rm sfr}\,\,}-2m_{\pi})\,\frac{1}{2q}
\arctan{\frac{q (M_{{\rm sfr}\,\,} - 2 m_{\pi})}{q^2 + 2 M_{{\rm sfr}\,\,} m_{\pi}}}\,.
\eea

\end{widetext}
Note that in the notations of Equations.~\ref{eq:EFTdecomp} and \ref{eq:xEFTdecomp}, the above finite range interactions do not contribute to the $\widetilde{V}_i$ and $\widetilde{V}^x_i$ interactions where $i \in C,S,T$.
For the numerical results presented in the text, we have assumed an axial-vector coupling of $g_A = 1.29$, a pion-decay constant of $f_{\pi}= 92.4$ MeV, and a pion mass of $m_{\pi} = 138$ MeV.  The $\pi\pi NN$ low energy constants have been extracted from  both $\pi N$ scattering~\cite{Bernard:1996gq} and NN phase shift analyses~\cite{Rentmeester:2003mf}, although the $c_3$ and $c_4$ couplings have rather large uncertainties whose central values are still somewhat controversial~\cite{Entem:2003cs}. In the results presented here, we have adopted the values $c_1 = -0.81$ GeV$^{-1}$, $c_3 = -3.4$ GeV$^{-1}$, and $c_4 = 3.4$ GeV$^{-1}$ as in Ref.~\cite{Epelbaum:2004fk}.

In Ref.~\cite{Epelbaum:2003gr}, Epelbaum and collaborators advocate using a spectral function regulator (SFR) mass in the range of $M_{{\rm sfr}\,\,} = 500\ldots 800$ MeV to reduce the unphysically strong attraction in the isoscalar central part of the N$^2$LO two-pion exchange potential (TPEP). In contrast, Entem and Machleidt work in a scheme that uses dimensional regularization to regulate the divergent loop integrals that enter into the calculation of the TPEP, which corresponds to taking $M_{{\rm sfr}\,\,} = \infty$~\cite{Entem:2003ft}.  As emphasized in~\cite{Epelbaum:2003gr}, formally speaking there is no ``correct'' value since varying $M_{{\rm sfr}\,\,}$ only modifies the short-distance structure of the TPEP, and such variations have no effect on observables since they can always be absorbed by the corresponding contact interactions. Nevertheless, it is argued in Ref.~\cite{Epelbaum:2003gr} that a finite SFR mass in the range of $M_{{\rm sfr}\,\,} = 500\ldots 800$ MeV offers certain practical advantages by cutting down the unnaturally strong attraction at mid- and short-distances that arises at N$^2$LO. For this reason, we have used a value of $M_{{\rm sfr}\,\,}=500$ MeV to generate the DME couplings shown in the figures in the text.

On a technical note, we mention that we were not able to obtain analytic expressions for the NLO couplings for finite values of $M_{{\rm sfr}\,\,}$. However, numerical integration of the ``master formulas'' Eqs.~\ref{eq:master1}-\ref{eq:master2} showed rather small differences from the $M_{{\rm sfr}\,\,}=\infty$ NLO couplings. For the N$^2$LO couplings where the SFR mass has a much more quantitative effect, we were able to obtain analytic expressions provided $M_{{\rm sfr}\,\,}$ is chosen to be an integer multiple of $m_{\pi}$. For all other choices, such as the $M_{{\rm sfr}\,\,} = 500$ MeV used in the text, the NLO and N$^2$LO DME couplings were obtained by numerical integration of Eqs.~\ref{eq:master1}-\ref{eq:master2}.

\section{Contact term EDF contributions}
\label{appendix:contacts}
Through N$^2$LO, the EFT two-nucleon contact interaction takes the form

\begin{widetext}
\bea
\label{Vcon}
\langle \pvec|V^{\text{NN}}_{ct}|\pvec'\rangle &=& C_S  + C_T \, \vec{\sigma}_1 \cdot  \vec{\sigma}_2 \,+\, C_1 \, \vec{q}\,^2 + C_2 \, \vec{k}^2 +
( C_3 \, \vec{q}\,^2 + C_4 \, \vec{k}^2 ) ( \vec{\sigma}_1 \cdot \vec{\sigma}_2)
+ iC_5\, \frac{1}{2} \, ( \vec{\sigma}_1 + \vec{\sigma}_2) \cdot ( \vec{q} \times
\vec{k})\nonumber\\
&& + C_6 \, (\vec{q}\cdot \vec{\sigma}_1 )(\vec{q}\cdot \vec{\sigma}_2 )
+ C_7 \, (\vec{k}\cdot \vec{\sigma}_1 )(\vec{k}\cdot \vec{\sigma}_2 )~,
\eea
\end{widetext}
where the $\Lambda$-dependence of the couplings has been suppressed for brevity.  Decomposing  $V^{\text{NN}}_{ct}$ in terms of the components $\{V_C, V_S,\ldots\}$ and $\{\widetilde{V}_C,\widetilde{V}_S,\ldots\}$ introduced in Eq.~\ref{eq:EFTdecomp} gives

\bea
V_C &=& C_S + C_1\qvec^2 \\
V_S &=& C_T + C_3\qvec^2 \\
V_T &=& C_6 \\
V_{LS} &=&  C_5
\eea
and
\bea
\widetilde{V}_C &=& C_2\kvec^2 \\
\widetilde{V}_S &=& C_4\kvec^2 \\
\widetilde{V}_T &=& C_7\,.
\eea

The EDF couplings arising from Eq.~\ref{Vcon} treated at the Hartree-Fock level are given by
\bea
\label{eq:contactcouplings}
&&C_0^{\rho\rho} = \frac{3}{8}(C_S - C_T) \\
&&C_1^{\rho\rho} = -\frac{1}{8}(C_S + 3C_T) \\
&&C_0^{\rho\tau} = \frac{1}{4}(C_2 - C_1 -3C_3-C_6)\\
&&C_1^{\rho\tau} = -\frac{1}{4}(C_1 + 3C_3 + C_6) \\
&&C_0^{\rho\Delta\rho} = \frac{1}{64}(C_2 - 16C_1 + 3C_4 + C_7) \\
&&C_1^{\rho\Delta\rho} = \frac{1}{64}(C_2 + 3C_4 +C_7) \\
&&C_0^{JJ} = \frac{1}{16}(2C_1 - 2C_3  -2C_4 -4C_6 + C_7) \\
&&C_1^{JJ} = \frac{1}{8}(C_1 - C_3 - 2C_6)\,.
\label{eq:contactcouplings2}
\eea
As discussed in the text, the Hartree-Fock energy density resulting from $V^{\text{NN}}_{ct}$ is precisely of the same form as empirical Skyrme functionals, i.e., bilinear products of densities multiplied by the density-independent coupling constants in Eqs.~\ref{eq:contactcouplings}-\ref{eq:contactcouplings2}. This observation motivates the semi-empirical approach advocated in the text whereby the density-dependent DME Fock couplings from the finite-range pion exchange contributions are added to existing Skyrme functionals, whose empirical parameters are then refit to data. In this sense, the refitted Skyrme constants can be viewed as containing the contributions of Eqs.~\ref{eq:contactcouplings}-\ref{eq:contactcouplings2} in addition to higher order contributions beyond HF.

\section{Exchange interaction}
\label{appendix:exchangeNN}
By evaluating the action of the exchange operators on the direct NN interaction in Eq.~\ref{eq:EFTdecomp}, the form factors of the exchange interaction $\{V^x_C,V^x_S,\ldots\}$ appearing in Eq.~\ref{eq:xEFTdecomp} can be expressed as linear combinations of the direct interaction $\{V_C,V_S,\ldots\}$ as shown in Table~\ref{tab:exchangeNN}.

\begin{table}[htdp]
\label{tab:exchangeNN}
\begin{tabular}{c||c|c|c|c|c|c|c|c}
 & $V_c$ & $W_c$ & $V_s$ & $W_s$ & $V_T$
 & $W_T$ & $V_{LS}$ & $W_{LS}$  \\
\hline
\hline
 $V^x_c$ & $\frac{1}{4}$ & $\frac{3}{4}$ & $\frac{3}{4}$ & $\frac{9}{4}$ & $k^2$
 &$3k^2$ & 0 & 0  \\
 \hline
 $W^x_c$ & $\frac{1}{4}$ & $-\frac{1}{4}$ & $\frac{3}{4}$ & $-\frac{3}{4}$ &
$k^2$ & $-k^2$ & 0 & 0  \\
 \hline
 $V^x_s$ & $\frac{1}{4}$ & $\frac{3}{4}$ & $-\frac{1}{4}$ & $-\frac{3}{4}$ &
 $-k^2$ & $-3k^2$ & 0 & 0 \\
 \hline
 $W^x_s$ & $\frac{1}{4}$ & $-\frac{1}{4}$ & $-\frac{1}{4} $ & $\frac{1}{4}$ &
 $-k^2$ & $k^2$ & 0 & 0\\
 \hline
 $V^x_T$ &0&0&0& 0& $\frac{1}{2}$ & $\frac{3}{2}$ & 0&0 \\
 \hline
 $W^x_T$ &0&0&0&0& $\frac{1}{2}$ &$-\frac{1}{2}$ & 0&0\\
 \hline
 $V^x_{LS}$ &0&0&0&0&0&0&$-\frac{1}{2}$ &$-\frac{3}{2}$  \\
 \hline
  $W^x_{LS}$ &0&0&0&0&0&0&$-\frac{1}{2}$ &$\frac{1}{2}$
\end{tabular}
\caption{Recoupling coefficients between the direct and exchange NN interaction. Note that the $\{V_C,V_S,\ldots\}$ are evaluated at $2\kvec$, since exchanging the two nucleons corresponds to replacing $\qvec \rightarrow 2\kvec$ and $\kvec \rightarrow \qvec/2$.}
\end{table}

\section{PSA-DME}

\label{appendix:psadme}
\begin{widetext}
In this appendix we provide a streamlined derivation of the PSA-DME compared to the original presentation in Ref.~\cite{Gebremariam:2009ff}, along with some details concerning the simplified version used in the present work.
All variants of the PSA-DME start from the formal identity for the scalar/vector-isoscalar/isovector part of the one-body density matrix
\begin{eqnarray}\label{eq:rhomunu}
\rho_{\mu \nu}(\vec{r}_1, \vec{r}_2)&=& e^{i \vec{r} \cdot \vec{k}}  e^{\vec{r} \cdot \bigl[ \frac{\vec{\nabla}_1-\vec{\nabla}_1}{2} - i \vec{k}\bigr]}
\, \sum_{i=1}^{A} \varphi^{\ast}_i (\vec{r}_2 \vec{\sigma}_2 \vec{\tau}_2)\,\varphi_i (\vec{r}_1 \vec{\sigma}_1 \vec{\tau}_1) \,
\langle \vec{\sigma}_2|\hat{\sigma}_{\mu}| \vec{\sigma}_1 \rangle \,\langle \vec{\tau}_2|\hat{\tau}_{\nu}| \vec{\tau}_1 \rangle \,\nonumber\\
&\approx &
e^{i \vec{r} \cdot \vec{k}}  \biggl[ 1 + \vec{r} \cdot \biggl(\frac{\vec{\nabla}_1 -\vec{\nabla}_2}{2} - i \vec{k}\biggr) +
\frac{1}{2} \biggl( \vec{r} \cdot \bigl(\frac{\vec{\nabla}_1 -\vec{\nabla}_2}{2} - i \vec{k}\bigr) \biggr)^2 \biggr]
\, \sum_{i=1}^{A} \varphi^{\ast}_i (\vec{r}_2 \vec{\sigma}_2 \vec{\tau}_2)\,\varphi_i (\vec{r}_1 \vec{\sigma}_1 \vec{\tau}_1)\nonumber\\
&&\qquad \,\times \,
\langle \vec{\sigma}_2|\hat{\sigma}_{\mu}| \vec{\sigma}_1 \rangle \,\langle \vec{\tau}_2|\hat{\tau}_{\nu}| \vec{\tau}_1 \rangle \,,\label{basicstartingeqn}
\end{eqnarray}
where indices $\mu, \, \nu \, \epsilon \, \{0,1,2,3\}$, $\tau_0$ and $\sigma_0$ correspond to
a two-by-two identity matrix, $\tau_{1,2,3} \, \equiv \, \tau_{x,y,z}$ and
$\sigma_{1,2,3} \, \equiv \, \sigma_{x,y,z}$. $\vec{k}$ is a yet-to-be-determined momentum scale to be chosen to optimize the truncated expansion in Eq.~\ref{basicstartingeqn}. Physically, $\vec{k}$ represents an averaged relative momentum in the nucleus. Now, assume we have a model local momentum distribution given by $g(\vec{R}, \vec{k})$ and define the following quantities
\begin{eqnarray}
\Pi_n (\vec{r}, \vec{R}) \,&\equiv&\, \frac{\int d \vec{k} \, e^{i \vec{r} \cdot \vec{k}} \bigl(\,\vec{r} \cdot \vec{k}\bigr)^n  g(\vec{R}, \vec{k})}{\int d \vec{k}
\, g(\vec{R}. \vec{k})}\,,\\
j_{a, \mu  \nu}(\vec{R}) \,&\equiv&\, -\frac{i}{2}\,\bigl(\vec{\nabla}^{(1)}_{a} - \vec{\nabla}^{(2)}_{a} \bigr)\,
\rho_{\mu \nu}(\vec{r}_1, \vec{r}_2) \,\biggl{|}_{\vec{r}_1 = \vec{r}_2 = \vec{R}}\,,\\
\tau_{a b, \mu \nu}(\vec{R}) \,&\equiv&\, \nabla^{(1)}_{a}\,\nabla^{(2)}_{b}\,
\rho_{\mu \nu}(\vec{r}_1, \vec{r}_2) \,\biggl{|}_{\vec{r}_1 = \vec{r}_2 = \vec{R}}\,.
\end{eqnarray}
Performing the phase-space averaging of Eq.~\eqref{basicstartingeqn} on the model space
defined by $g(\vec{R}, \vec{k})$, we obtain
\begin{eqnarray}
\rho_{\mu \nu}(\vec{r}_1, \vec{r}_2)
&\approx &
\biggl[ \Pi_0 + \Pi_0\, \vec{r} \cdot \frac{\vec{\nabla}_1 - \vec{\nabla}_2}{2} - i\, \Pi_1
+ \frac{\Pi_0}{2}\biggl( \vec{r} \cdot \frac{\vec{\nabla}_1 - \vec{\nabla}_2}{2}\biggr)^2
- \frac{\Pi_2}{2} - i \Pi_1\,
\biggl(\vec{r} \cdot \frac{\vec{\nabla}_1 - \vec{\nabla}_2}{2} \biggr)
\biggr]\nonumber\\
&&\qquad\times\,
\, \sum_{i=1}^{A} \varphi^{\ast}_i (\vec{r}_2 \vec{\sigma}_2 \vec{\tau}_2)\,\varphi_i (\vec{r}_1 \vec{\sigma}_1 \vec{\tau}_1) \,
\langle \vec{\sigma}_2|\hat{\sigma}_{\mu}| \vec{\sigma}_1 \rangle \,\langle \vec{\tau}_2|\hat{\tau}_{\nu}| \vec{\tau}_1 \rangle \,
\bigg{|}_{\vec{r}_1 = \vec{r}_2 = \vec{R}}\,,\nonumber\\
&\approx& \biggl[\Pi_0 - i \Pi_1 - \frac{\Pi_2}{2}\biggr] \rho_{\mu \nu}(\vec{R})
\, + \, i \biggl[\Pi_0 - i\, \Pi_1\biggr] \sum_{a} \,r_a\, j_{\mu a k}(\vec{R})
\,\nonumber\\
 && \qquad + \,  \frac{\Pi_0}{2}\,\sum_{a, b}\,r_a \, r_b\, \biggl[ \frac{1}{4}\nabla_a\nabla_b\rho_{\mu \nu}(\vec{R}) - \tau_{a b, \mu \nu}(\vec{R})\biggr]\,,
\label{tempresult1}
\end{eqnarray}
where the local densities are as defined previously.

Even without specifying
the actual form of the model momentum distribution, it is clear that the PSA-DME of the scalar and vector parts are treated on equal footing (i.e., $\Pi^{\rho}_n = \Pi^{\vec{s}}_n$). As shown in Ref.~\cite{Dobaczewski:2010qp}, elementary constraints derived from spin-polarized infinite matter forbid the use of channel-dependent $\Pi$-functions. Unfortunately, the channel-independence of the PSA-DME  $\Pi$-functions is not entirely transparent from the presentation in Ref.~\cite{Gebremariam:2009ff} because (i) we only considered the time-reversal invariant case such that only $\Pi^{\rho}_0$ and $\Pi^{\rho}_2$ had to be dealt with for the scalar part while only $\Pi^{\vec{s}}_1$ had to be dealt with for the vector part and (ii) an unnecessary asymmetry was introduced in the form of an additional angle-average over the direction of $\vec{r}$ for the scalar part in order to replace the kinetic tensor density $\tau_{ab}$ with the diagonal kinetic density $\tau$. The less transparent derivations in Ref.~\cite{Gebremariam:2009ff} obscure the fact that intrinsically, the PSA leads to a channel-independent DME with the same $\Pi$-functions for the scalar and the vector parts. In view of the simplified (and more general)  derivation presented in the current paper, it should be realized that
the claim in Ref.~\cite{Dobaczewski:2010qp} that the PSA-DME postulates different $\Pi$-functions for the scalar and vector parts is no longer correct.

As discussed in Ref.~\cite{Gebremariam:2009ff}, the PSA-DME is well-suited to incorporate the effects of the diffuseness and anisotropy of the local momentum distribution at the spatial surface. However, the inclusion of the diffuseness complicates calculations since analytical expressions can no longer be obtained, thus introducing the need for fit parameters into the formalism if one desires analytical parameterizations for the couplings.  Since the diffuseness primarily affects the expansion of the scalar part, which is already reasonably accurate in all existing variants of the DME, we neglect it here for simplicity. On the other hand, the inclusion of the anisotropy of the local momentum distribution at the nuclear surface predominantly affects the DME of the vector part.  As shown in Ref.~\cite{Gebremariam:2009ff}, the anisotropy does not modify the functional form of the $\Pi^{\vec{s}}_1$ function, as it only enters into the definition of the local Fermi momentum, see Eqs.~\ref{eq:kFdeformed} and \ref{eqn:P_2definition}. This does not complicate the evaluation of the energy, but it does result in significantly more complicated single-particle fields in the self-consistency loop. Therefore, we take a simplified approach (which still gives substantial improvements over the original NV-DME for the vector part) by using the phase space of symmetric nuclear matter to perform the averaging, i.e. $g(\vec{R}, \vec{k}) = \Theta (k_F - \vec{k})$. As a result, we find
\begin{eqnarray}
\Pi_0 (k_F r) \, &=&\, 3 \,\frac{j_1 (k_F r)}{k_F r} \,\approx \, 1 \,+\, {\cal O}(k_F r)^2\,,\\
\Pi_1 (k_F r) \, &=&\, -\,i\, 3 \,j_0 (k_F r) \, + \,i \,9\,\frac{j_1 (k_F r)}{k_F r} \, \approx\,
i\,\frac{(k_F r)^2}{5} \,+\, i \,{\cal O}(k_F r)^4\,,\label{Pi1}\\
\Pi_2 (k_F r) \, &=&\, 15 \,j_0 (k_F r) \, - \, 36 \frac{j_1 (k_F r)}{k_F r}  \,-\,3\,\text{cos}(k_F r)\,\approx\,
\frac{(k_F r)^2}{5}\, +\, {\cal O}(k_F r)^4\,.
\end{eqnarray}
While $\Pi_0$ starts from $1$, the other two $\Pi-$functions start from
$ {\cal O}(k_F r)^2$. Using a weak ordering argument that counts $k_F$ on the same ground as the number of gradients,
Eq.~\eqref{tempresult1} can be rearranged as
\begin{eqnarray}
\rho_{\mu \nu}(\vec{r}_1, \vec{r}_2)
&\approx& \Pi_0  \rho_{\mu \nu}(\vec{R})
\, + \, i \Pi_0 r_a j_{a, \mu  \nu}(\vec{R})
\, + \,  \frac{\Pi_0}{2}\,r_a \, r_b\,\biggl[
 \frac{1}{4}\nabla_a\nabla_b\rho_{\mu \nu}(\vec{R}) - \tau_{a b, \mu \nu}(\vec{R})\,
+ \frac{ \delta_{ab}\,\Lambda (k_F r) \, k^2_F}{5} \rho_{\mu \nu}(\vec{R})\biggr]\,,
\end{eqnarray}
where we neglected $i\, \Pi_1\,\sum_{a}\, r_a\, j_{\mu a k}(\vec{R})$ that turns out to be
a third-order correction (one gradient in the density and $k^2_F$ in $\Pi_1(k_F r)$).
The prefactor $\Lambda (k_F r)$ is defined as
\begin{eqnarray}
\Lambda (k_F r) \, \equiv \, -5 \, \frac{i\, 2 \, \Pi_1(k_F r) \, +\, \Pi_2(k_F r) }{k^2_F\, r^2\, \Pi_0(k_F r)} \,\approx\,1 + {\cal O}(k_F r)^2 \,.\label{lambda}
\end{eqnarray}
Approximating $\Lambda (k_F r) \approx 1 $, one recovers Eq. 44 of Ref.~\cite{Dobaczewski:2010qp}
as
\begin{eqnarray}\label{eq:rhomunudme}
\rho_{\mu \nu}(\vec{r}_1, \vec{r}_2)
&\approx& \Pi_0\,  \rho_{\mu \nu}(\vec{R})
\, + \, i\, \Pi_0 \,\sum_{a}\,r_a\, j_{a, \mu  \nu}(\vec{R})
\, + \,  \frac{\Pi_0}{2}\,\sum_{a, b}\,r_a \, r_b\,\biggl[
 \frac{1}{4}\nabla_a\nabla_b\rho_{\mu \nu}(\vec{R}) - \tau_{ a b, \mu \nu}(\vec{R})\,\nonumber\\
 &&\qquad + \delta_{ab} \frac{ \, k^2_F}{5} \rho_{\mu \nu}(\vec{R})\biggr]\,.
\label{finalresult1}
\end{eqnarray}
Note that within this approximation scheme (i.e., using the weak $k_F$-ordering and approximating $\Lambda(k_Fr)\approx 1$), the constraints on the $\Pi-$function resulting from requiring gauge invariance of the energy
density functional are satisfied trivially.

In order to recover Eqs.~\ref{eq:DMEscalar} and \ref{eq:DMEvector}, one first uses the fact that
the one-body density matrix is taken to be diagonal in isospin space. This implies
that $\rho_{\mu \nu}(\vec{r}_1, \vec{r}_2)$ is non-zero only if $\nu = \{0, 3\}$. Hence,
we identify the correspondence
\begin{eqnarray}\label{eq:rhomunucorrespondence}
\rho_{00}\, &=&\,\rho_{0} (\vec{r}_1, \vec{r}_2) \,\nonumber\\
\rho_{03}\, &=&\,\rho_{1} (\vec{r}_1, \vec{r}_2) \,\nonumber\\
\rho_{i 0}\, &=&\,s_{0, i} (\vec{r}_1, \vec{r}_2) \,\nonumber\\
\rho_{i 3}\, &=&\,s_{1, i} (\vec{r}_1, \vec{r}_2) \,,
\end{eqnarray}
where $i \,\epsilon\, \{1,2,3\}$. Starting with Eq.~\eqref{eq:rhomunucorrespondence}
one can obtain the corresponding relations involving the local densities. For instance,
$j_{a, 00}(\vec{R})$ is the current density $j_{0, a} (\vec{R})$, while $j_{a, i 0}$ is the
tensor spin-orbit density $J_{0, a i}(\vec{R})$. To obtain Eq.~\eqref{eq:DMEscalar}, we
perform angle averaging over the orientation of $\vec{r}$. Using the identity
\begin{equation}
\frac{1}{4 \pi} \int d \vec{e}_r \, (\vec{r} \cdot \vec{A})
(\vec{r} \cdot \vec{B}) = \frac{ r^2}{3} \vec{A} \cdot \vec{B}\,,
\end{equation}
and noting that the current density $\vec{j}_{0/1}(\vec{R})$ vanishes in time-reversal invariant systems,
one obtains Eq.~\eqref{eq:DMEscalar} with $\Pi^{\rho}_2 = \Pi^{\rho}_0$. We reiterate that
unlike NV-DME, the $\Pi-$functions of PSA-DME satisfy the constraint from gauge invariance trivially.
As given, Eq.~\eqref{eq:DMEvector} corresponds to the DME of $\vec{s}_{t} (\vec{r}_1, \vec{r}_2)$
for spherical systems. To obtain that, one combines Eq.~\eqref{eq:rhomunudme} and Eqs.~\eqref{eq:rhomunucorrespondence} to give
\begin{equation}\label{basicvectorexpansion}
s_{t,\nu} \biggl(\vec{R}+ \frac{\vec{r}}{2} , \vec{R}-
\frac{\vec{r}}{2}\biggr)  \simeq   i \,
\Pi^{\vec{s}}_1 (k_F r) \,\sum^{z}_{\mu=x} r_{\mu} J_{t, \mu \nu} (\vec{R}) \,.
\end{equation}
Further reduction is possible in spherical systems,
for which one can write  $J_{t,\mu \nu} (\vec{R}) $ as a sum of
pseudoscalar, vector and (antisymmetric) traceless tensor parts
\begin{equation}
J_{t , \mu \nu} (\vec{R}) = \frac{1}{3} \, \delta_{\mu
\nu} \, J^{(0)}_{t} (\vec{R})\, + \, \frac{1}{2} \,
\sum^{z}_{k=x}\epsilon_{\mu \nu k}\, J^{(1)}_{t,k} (\vec{R}) \,+ \, J^{(2)}_{t ,\mu \nu} (\vec{R})\,,
\end{equation}
where the three components read
\begin{eqnarray}
J^{(0)}_{t} (\vec{R}) &\equiv& \sum^{z}_{\mu,\nu=x}\,\delta_{\mu \nu}\,
J_{t , \mu \nu} (\vec{R}) \, ,\\
J^{(1)}_{t,k} (\vec{R}) &\equiv& \sum^{z}_{\mu,\nu=x} \epsilon_{\mu \nu k} \, J_{t ,
\mu \nu} (\vec{R}) \, ,\\
J^{(2)}_{t ,\mu \nu} (\vec{R}) &\equiv& J_{t , \mu \nu} (\vec{R}) - \frac{1}{3} \, \delta_{\mu \nu} \, J^{(0)}_{t} (\vec{R}) \, -\, \frac{1}{2} \, \sum^{z}_{k=x}\epsilon_{\mu \nu
k}\, J^{(1)}_{t,k} (\vec{R}) \,.
\end{eqnarray}
Noting that only the vector part survives in spherical systems, one can recovers Eq.~\eqref{eq:DMEvector}
from Eq.~\eqref{basicvectorexpansion}.

\end{widetext}

\section{Single particle fields}
\label{appendix:spfields}
The density-dependence introduced by the DME of all couplings appearing in the EDF
makes the single-particle fields more complex than the usual case
where only the $C_t^{\rho\rho}$ couplings are density-dependent.
Note that one can generate density-independent couplings by setting the
DME momentum scale $k_F(\vec{r})$
equal to a constant, while setting $k_F(\vec{r}) = 0$ recovers the naive Taylor series expansion.

For a systematic comparison of the single-particle fields that result from
the density-dependent and density-independent cases, we consider two cases:
one where we keep the DME momentum scale intact and the other where
we set $k_F (\vec{r}) = 0$. Also, we consider a spherical system where
the single-particle equation of motion is given by
\begin{equation}
h^{\tau} \phi_i(\vec{r} \tau) \, =\, \epsilon_{i \tau}\, \varphi(\vec{r} \tau)\,,
\end{equation}
with the single-particle spinor
\begin{eqnarray}
\phi_i(\vec{r} \tau) \, &\equiv & \, \left(\begin{array}{c} \varphi_i (\vec{r} \, \sigma=+\frac{1}{2} \,\tau)
 \\ \varphi_i(\vec{r} \, \sigma=-\frac{1}{2} \, \tau) \end{array} \right)\,.
\end{eqnarray}
The structure of $h^{\tau}$ and its various components read
\begin{eqnarray}
h^{\tau} \, &=& \,- \vec{\nabla} \cdot B_{\tau} (r) \vec{\nabla} \, + \,
U_\tau (r) \, - \,i \vec{W}_\tau \cdot \vec{\nabla} \times \vec{\sigma}\,,\nonumber\\
B_\tau (r) \, &=&\, \frac{\delta E}{\delta \tau_\tau}\,,\nonumber\\
U_\tau (r) \, &=&\, \frac{\delta E}{\delta \rho_\tau}\,,\nonumber\\
\vec{W}_\tau (r) \, &=&\, \frac{\delta E}{\delta \vec{J}_\tau}\,,
\end{eqnarray}
where $E=E[\rho,\tau,J]$ is the Hartree energy plus the DME approximation to the Fock energy.
Since in this work the density-dependence of the couplings is encoded in $k_F(\vec{r})$, the only
systematic difference between the fields in the two cases appears in $U_\tau(\vec{r})$. I.e.
\begin{equation}
U_\tau(r) \, \equiv \, U^\rho_\tau (r) + U^{k_F}_\tau (r)\,,
\end{equation}
where $U^\rho_\tau (r)$ denotes the field that results after setting $k_F (r) = 0$ while
$U^{k_F}_\tau (r)$ is due to the explicit density-dependence of the couplings
\begin{eqnarray}
U^{\rho}_\tau (r) \, &\equiv& \,\frac{\delta E|_{k_F(r) =0}}{\delta \rho_\tau}\,,\\
U^{k_F}_\tau(r) \,&\equiv&\,\frac{\delta E}{\delta k_F} \frac{\delta k_F}{\delta \rho_\tau} \,.
\end{eqnarray}

\section{Hartree couplings}
\label{appendix:hartree}
As discussed in the text, it is desirable to treat the Hartree energy exactly since the DME is known to perform poorly for such contributions. Nevertheless, it is possible to apply the DME to reduce the finite-range Hartree energy to the form of a local EDF. For completeness, we provide expressions that can be used to calculate the Hartree EDF couplings. In the present case, the Hartree energy Eq.~\ref{eq:finalH} simplifies since the only non-zero finite-range NN contribution arises from the central force $\Gamma^t_c(r)$,

\begin{equation}
V_H = \frac{1}{2}\sum_{t=0,1}\int d\Rvec d\rvec \, \rho_t(\Rvec+\rvec/2)\rho_t(\Rvec-\rvec/2)\,\Gamma^t_C(\rvec)\,.
\label{eq:directE}
\end{equation}

Following Negele and Vautherin~\cite{negele72}, we take
\bea
&&\rho_t(\Rvec+\rvec/2)\rho_t(\Rvec-\rvec/2)\approx \rho^2_t(\Rvec) \\
&&+ \,\frac{1}{2}r^2g(k_Fr)\Bigl[ \rho_t(\Rvec)\nabla^2\rho_t(\Rvec)-|\nabla\rho_t(\Rvec)|^2\Bigr]\,\nonumber,
\label{eq:directDME}
\eea
where $g(x) = 35\,j_3(x)/2x^3$. Inserting Eq.~\ref{eq:directDME} into Eq.~\ref{eq:directE} gives
the following additional EDF couplings from the Hartree energy

\bea
&&\delta C^{\rho\rho}_t = \frac{1}{2}\,\Gamma^{t}_c(\qvec=0) \,,\\
&&\delta C^{\rho\Delta\rho}_t = \frac{1}{2\pi k_F^5}\, \int q^2dq \Gamma^{t}_c(q)I_5(q/k_F) \,,\label{hartreerhodeltarho}\\
&& \delta C^{(\nabla\rho)^2}_t = - \delta C^{\rho\Delta\rho}_t \,,
\eea
where
\bea
I_5(\bar{q}) &=& \int x^4dx g(x)j_0(\bar{q}x) \,,\\
&=& -\frac{35\pi}{8}\bigl(5\bar{q}^2 - 3\bigr)\,\theta(1 - \bar{q})\,.
\eea
In Ref.~\cite{Dobaczewski:2010qp}, the authors advocate using a pure Taylor series
expansion for the Hartree contribution. This amounts to setting
$g(x) =1 $, which results in
\begin{equation}
I_5(\bar{q}) \, = \,  \frac{\pi}{\bar{q}} \frac{d^3 \delta(\bar{q})}{d \bar{q}^3} \,.
\end{equation}
Using this value of $I_5(\bar{q})$ in Eq.~\eqref{hartreerhodeltarho}, one obtains the corresponding EDF
couplings from the Hartree energy.

\section{DME couplings from Mathematica}
\label{appendix:couplings}

In this section,we give explicit expressions for the 
EDF couplings that result from the application of NV-DME
and PSA-DME to the Fock energy from NN chiral EFT interaction thru
N$^2$LO. The couplings shown are calculated for $M_{{\rm sfr}\,\,} = \infty$ since the analytical expressions are more compact than for finite values of the SFR mass. For a more complete listing of the EDF couplings (including the capability to re-calculate the couplings using finite values of $M_{{\rm sfr}\,\,}$), refer to the Mathematica notebook that is provided with this submission. The symbol names for the couplings follows the
simple rule: DME-type + ``C'' + EDF-term + iso-scalar/iso-vector
+``noSFR''. The remaining Mathematica symbols are self-explanatory.

\afterpage{\clearpage\begin{widetext}
\begin{figure}[p]
\includegraphics{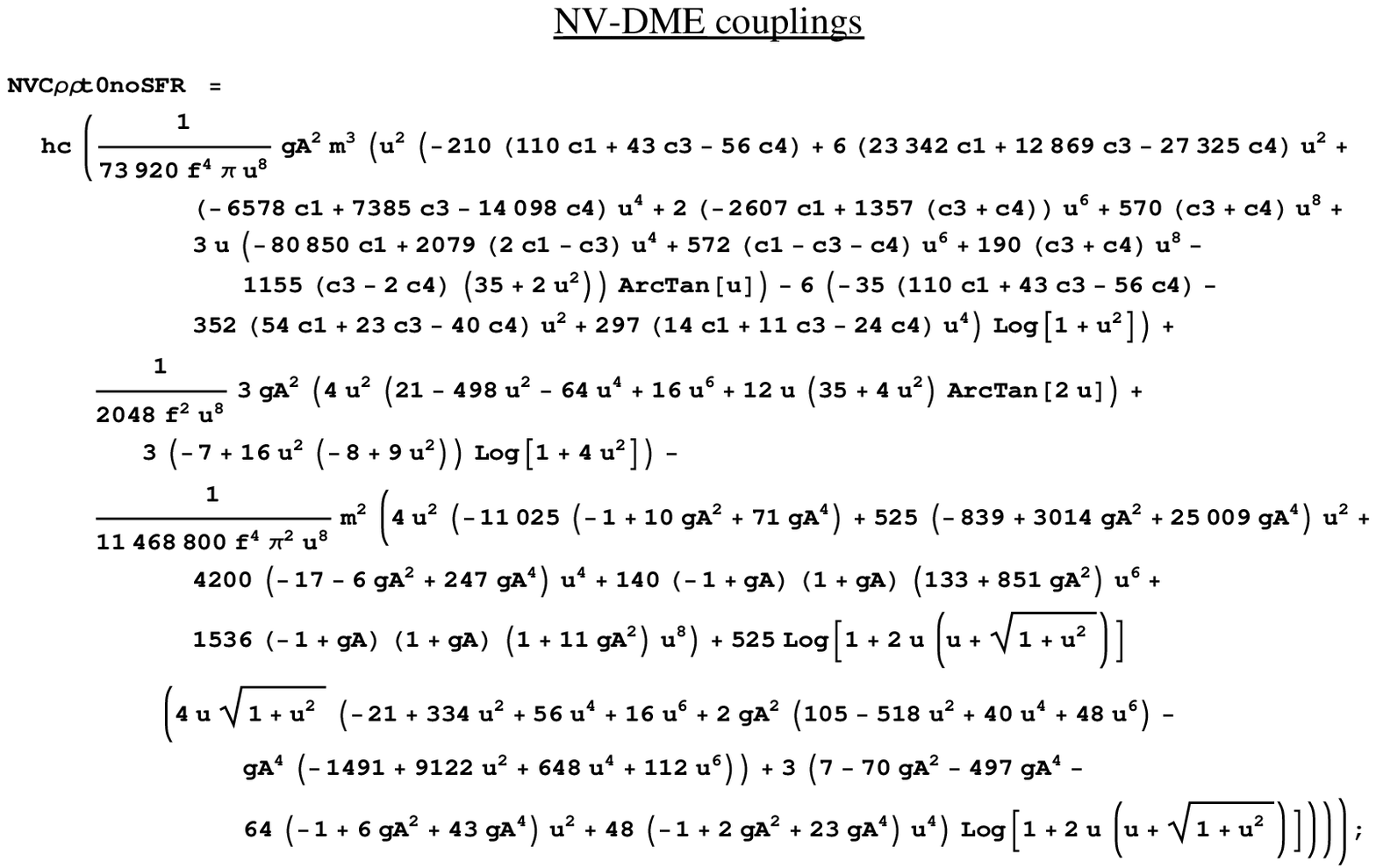}
\includegraphics{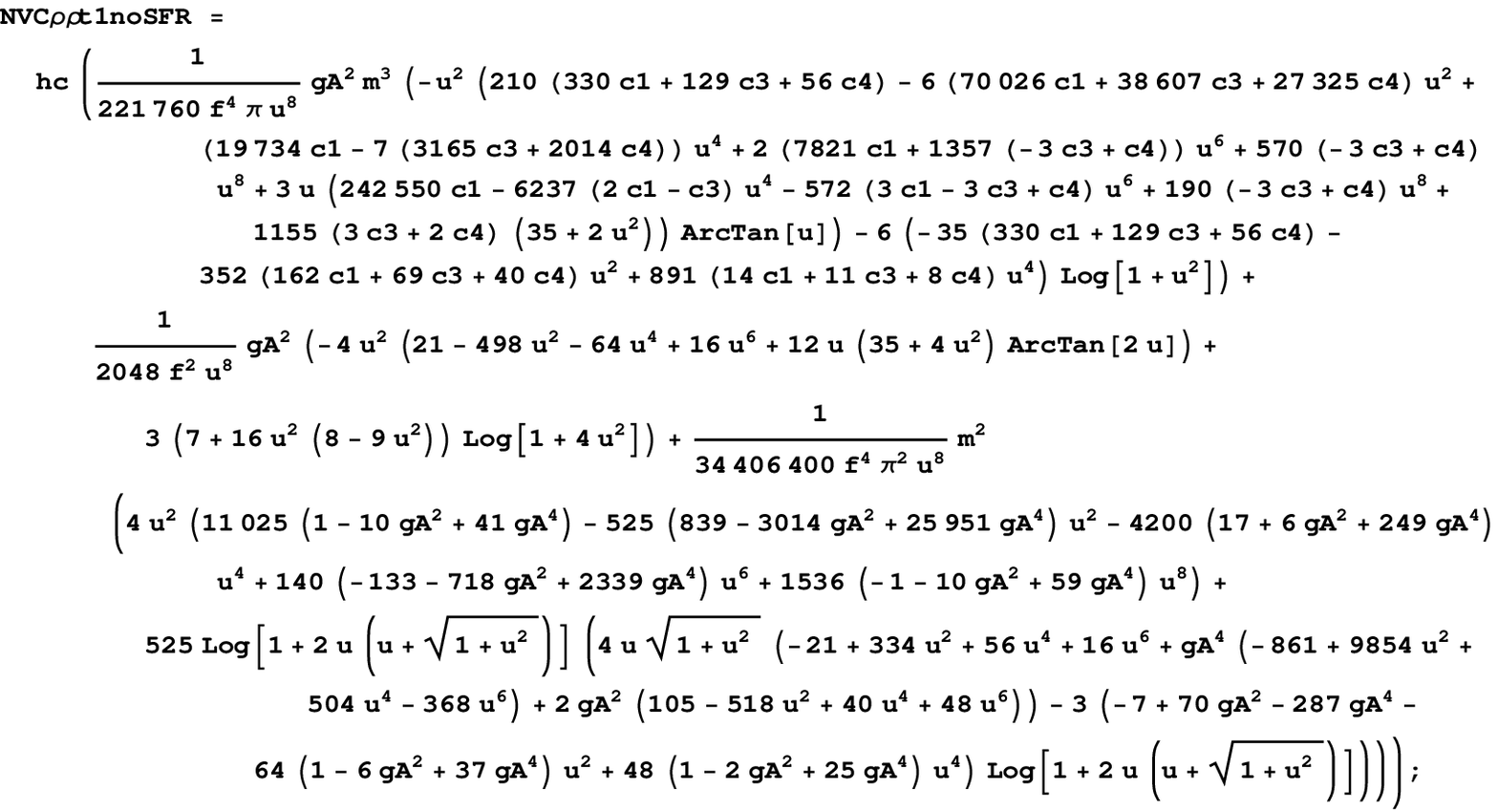}
\end{figure}

\begin{figure}[p]
\includegraphics{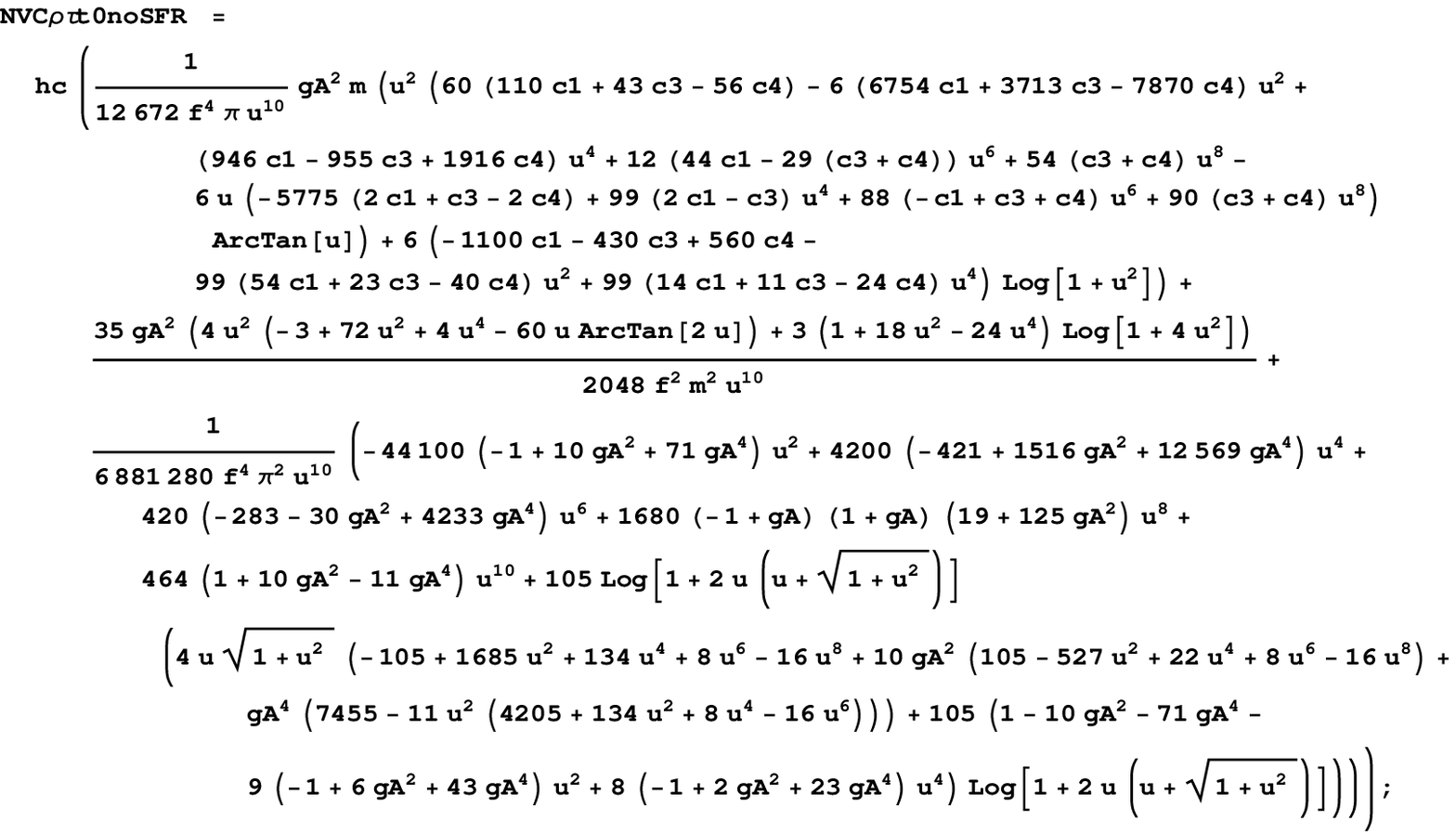}
%
\includegraphics{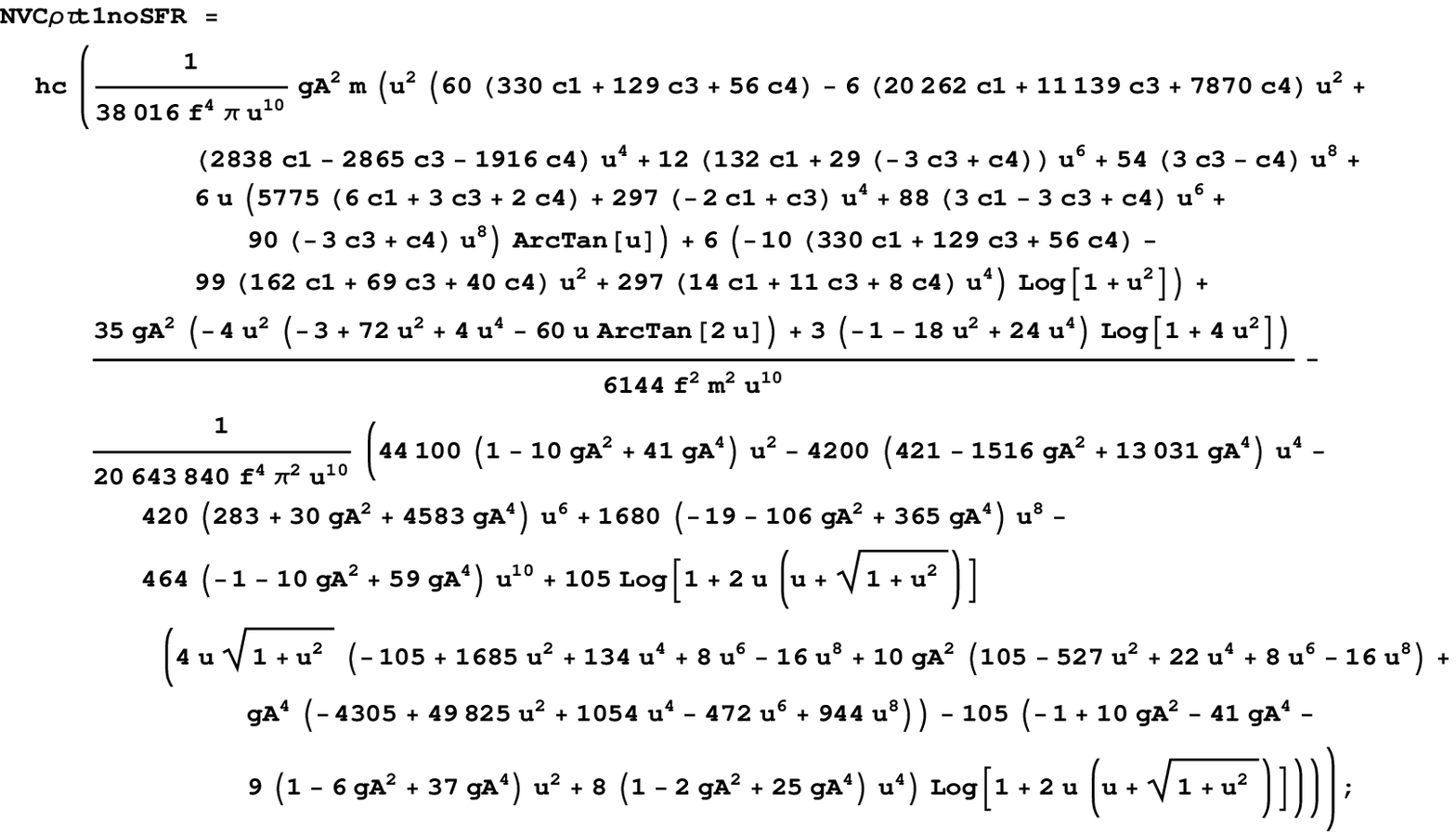}
\end{figure}

\begin{figure}[p]
\includegraphics{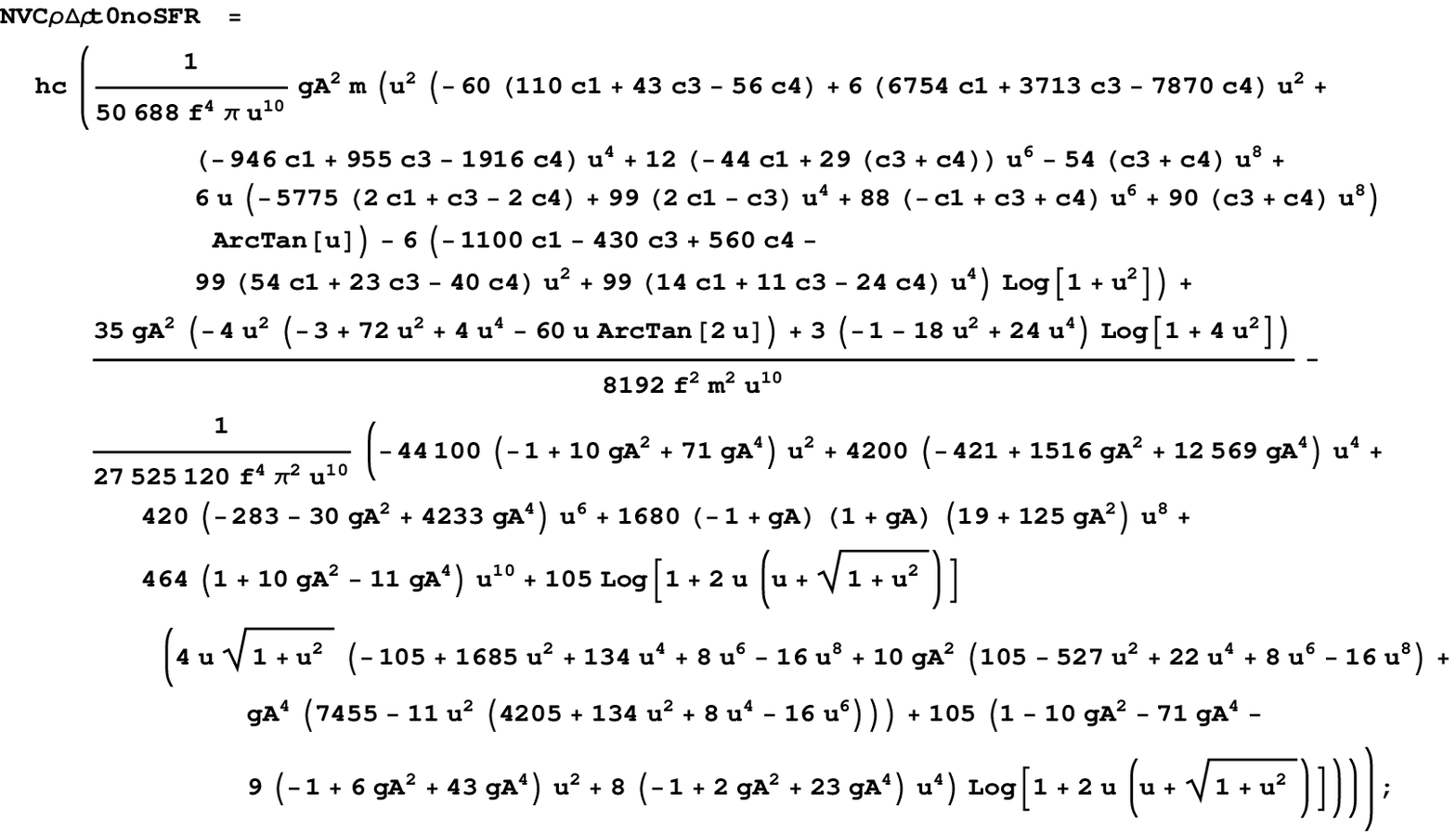}
%
\includegraphics{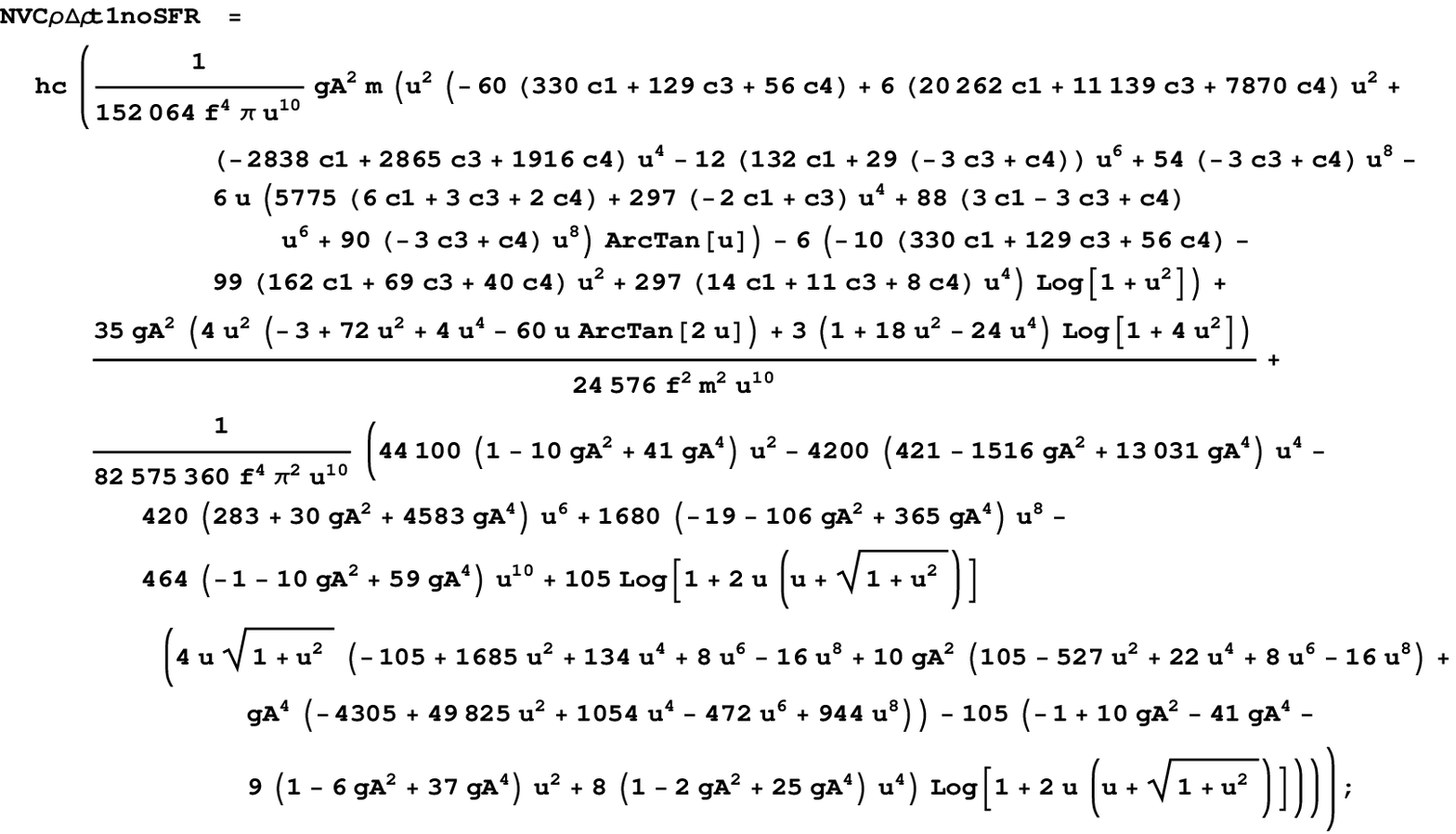}
\end{figure}

\begin{figure}[p]
\includegraphics{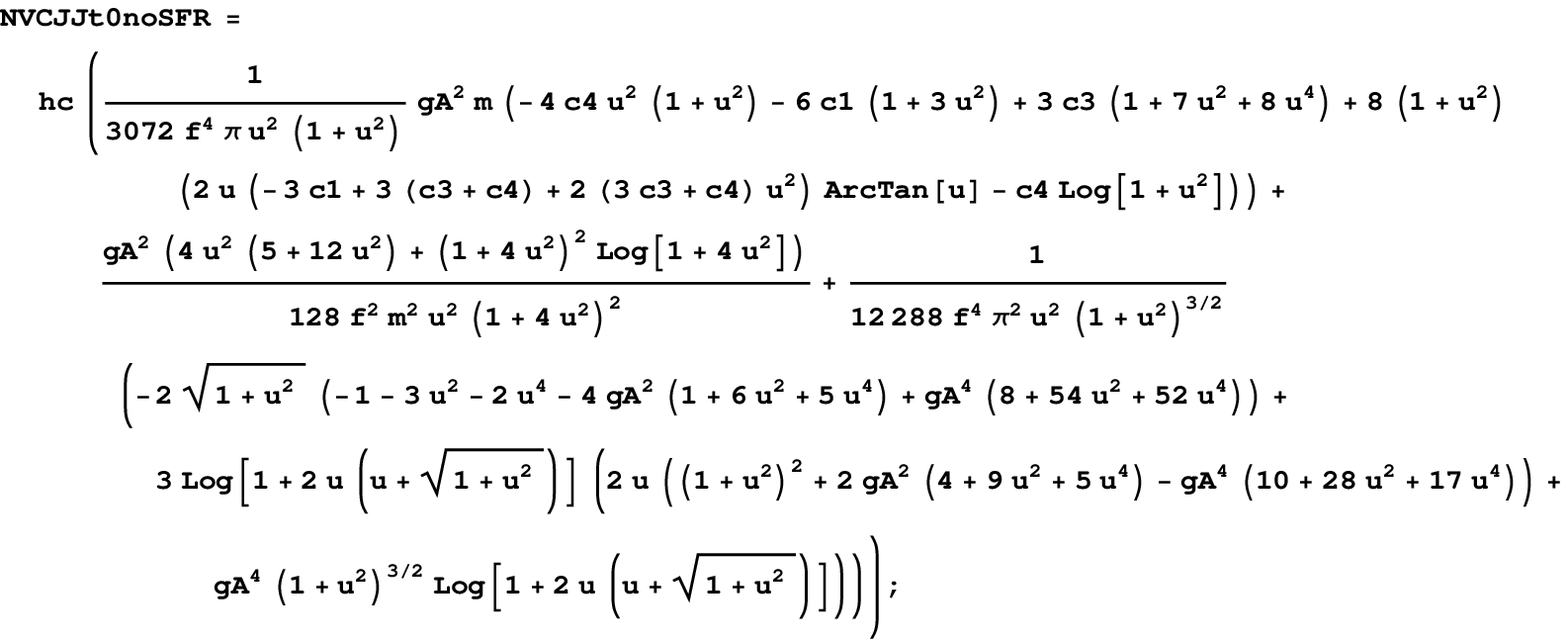}
%
\includegraphics{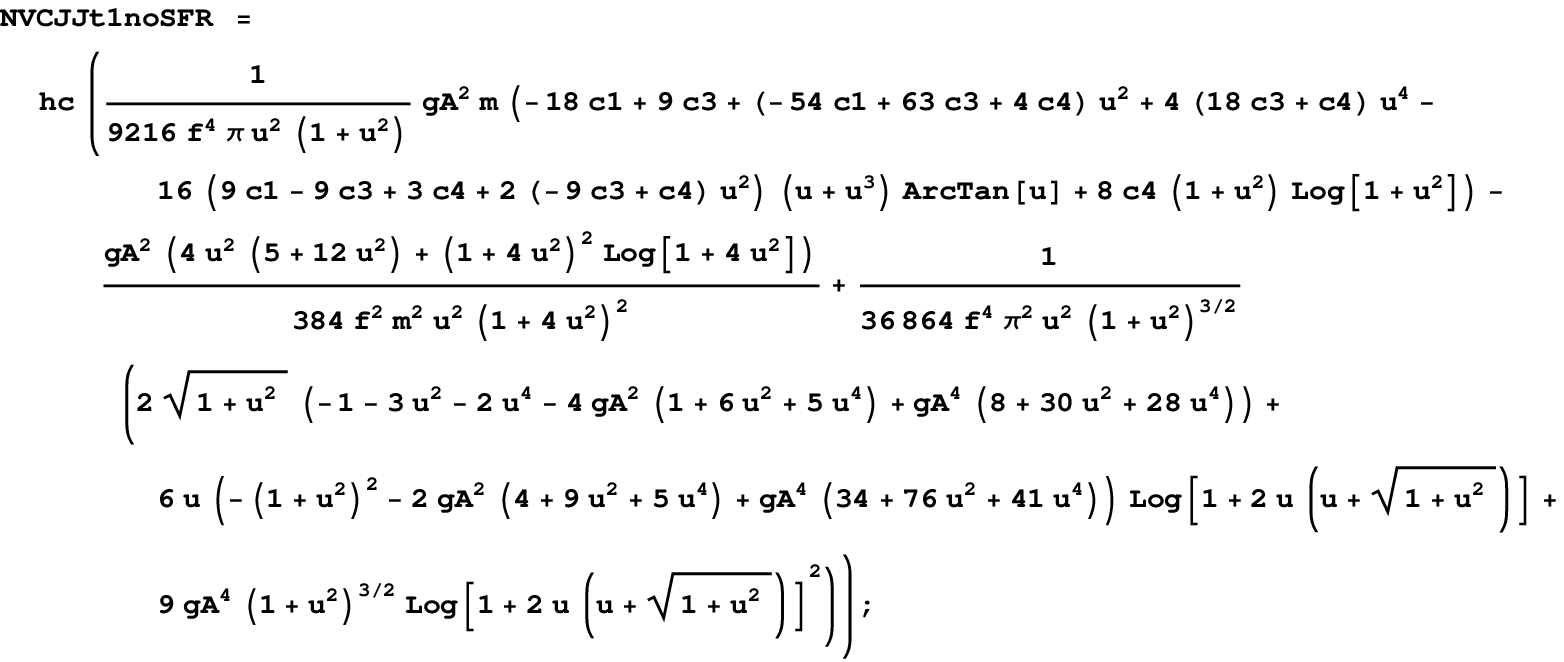}
\end{figure}
\end{widetext}

}
\afterpage{\clearpage\begin{widetext}
\begin{figure}[p]
\includegraphics{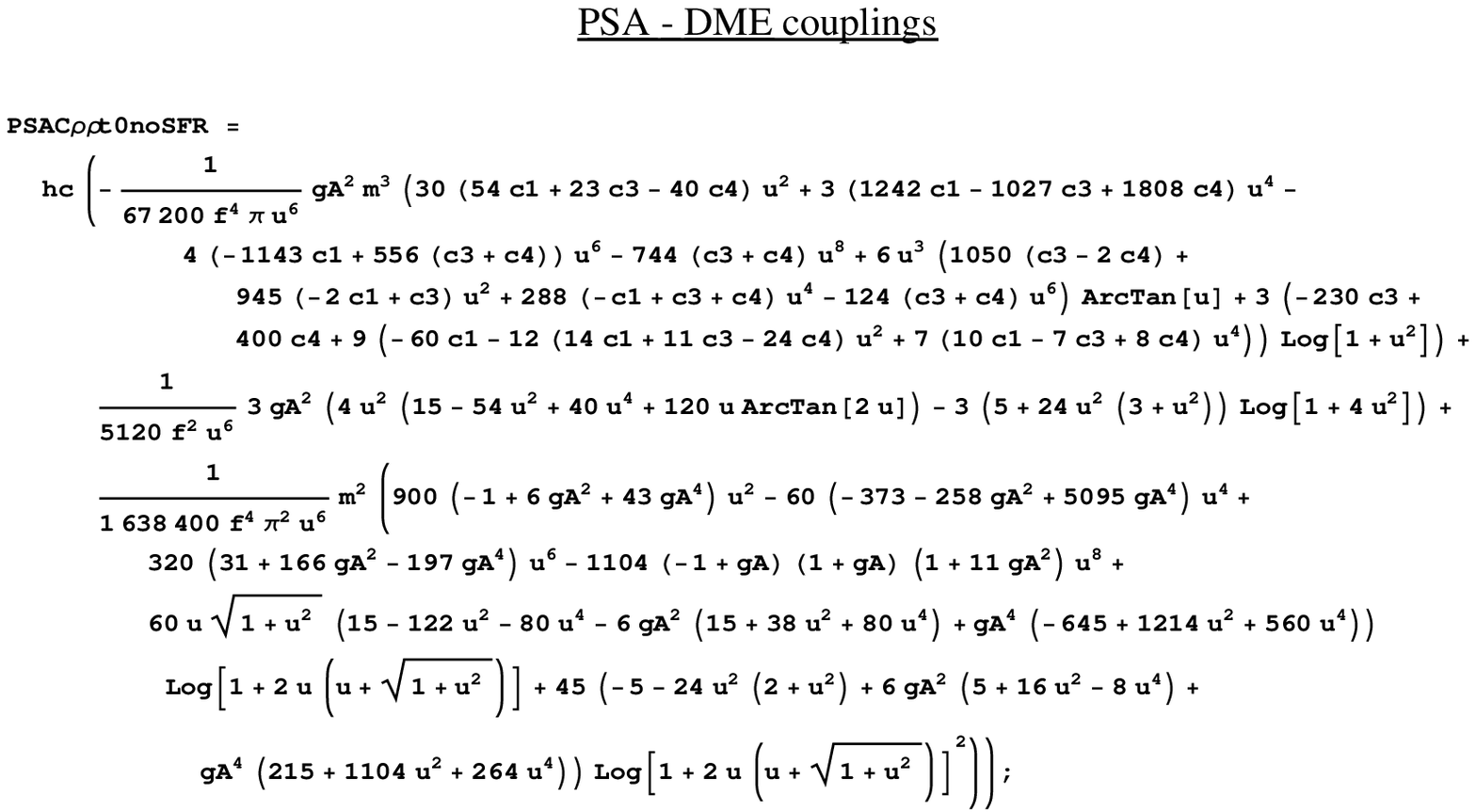}
%
\includegraphics{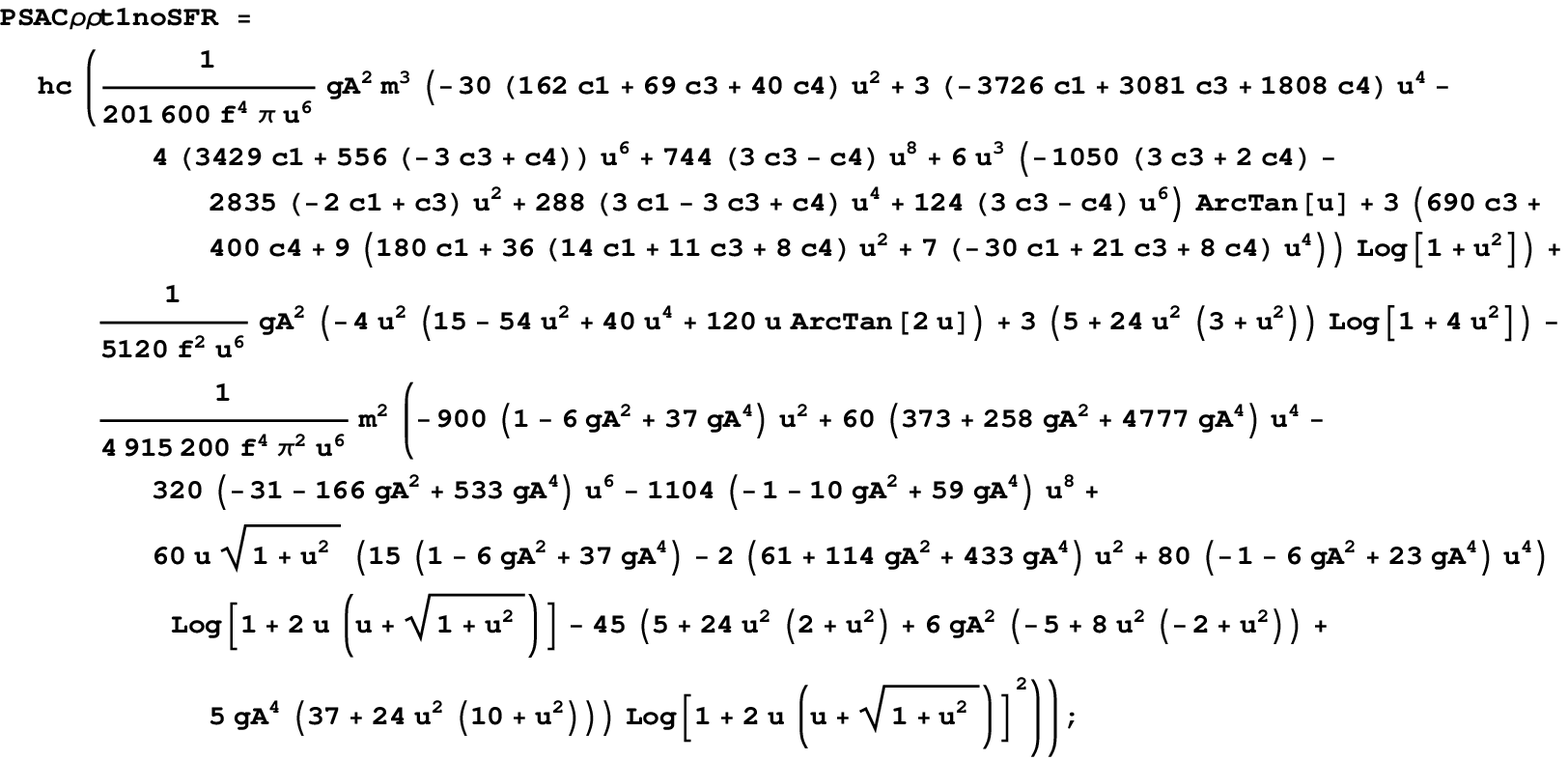}
%
\includegraphics{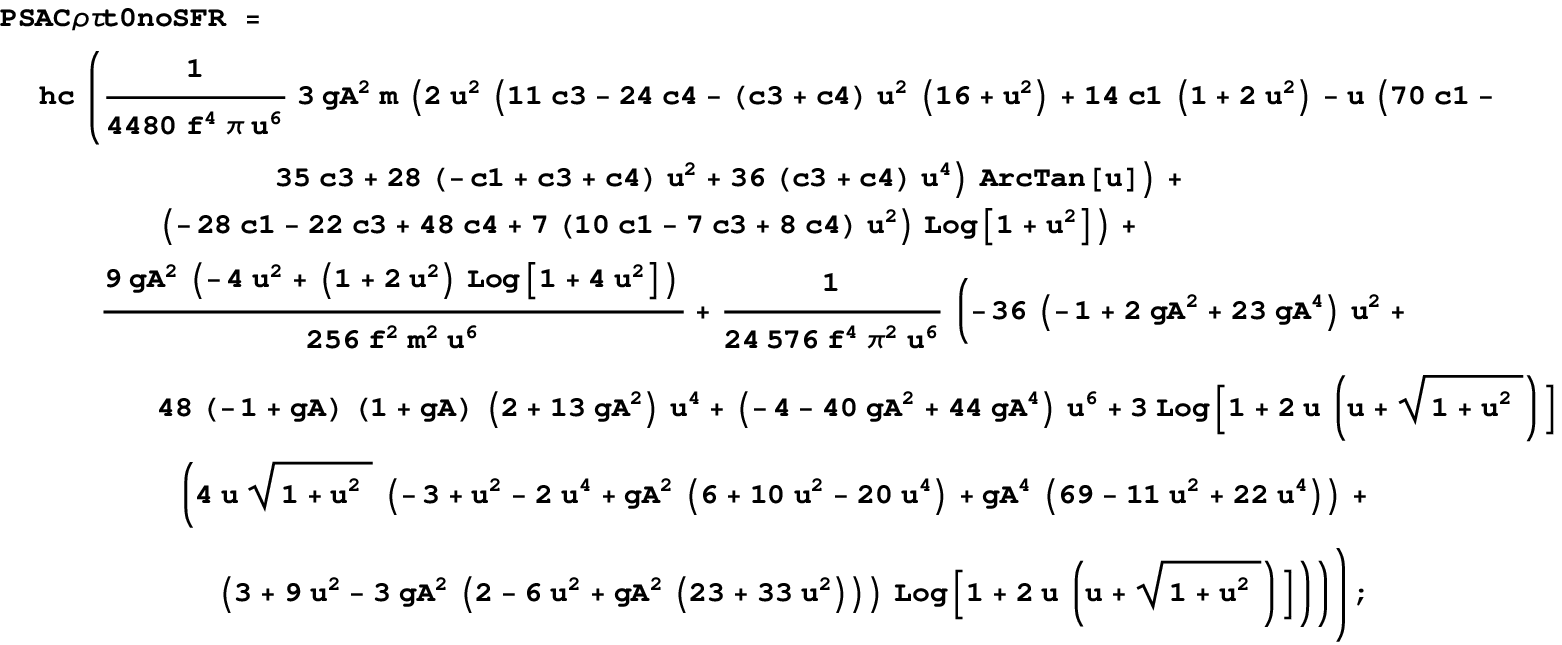}
\end{figure}

\begin{figure}[p]
\includegraphics{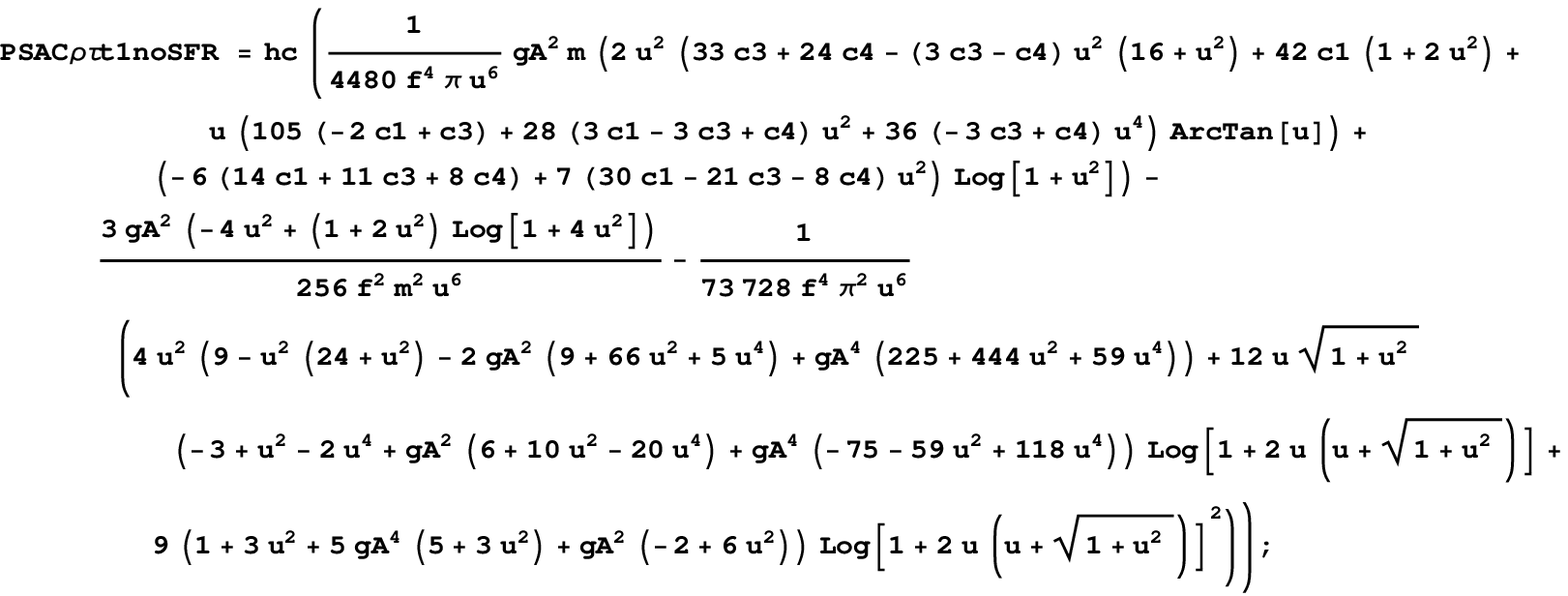}
\end{figure}

\begin{figure}[p]
\includegraphics{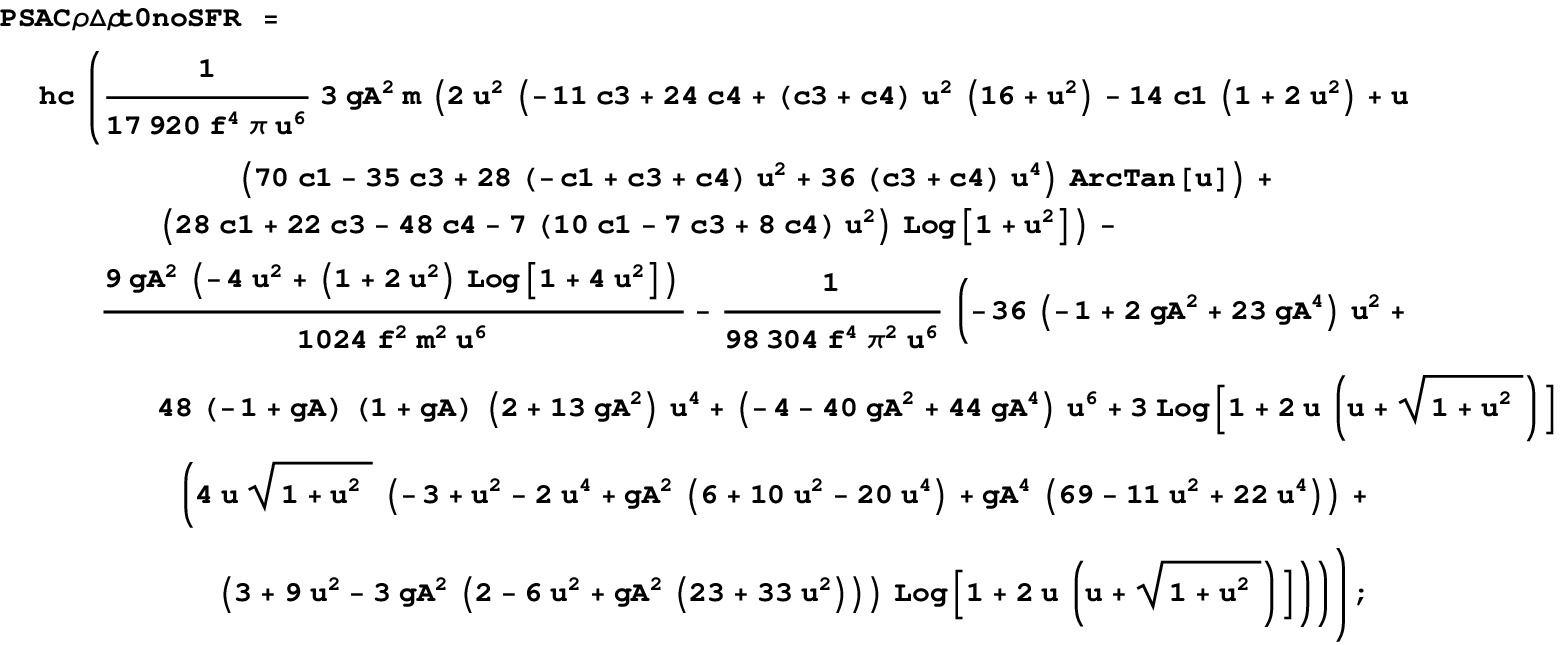}
\end{figure}

\begin{figure}[p]
\includegraphics{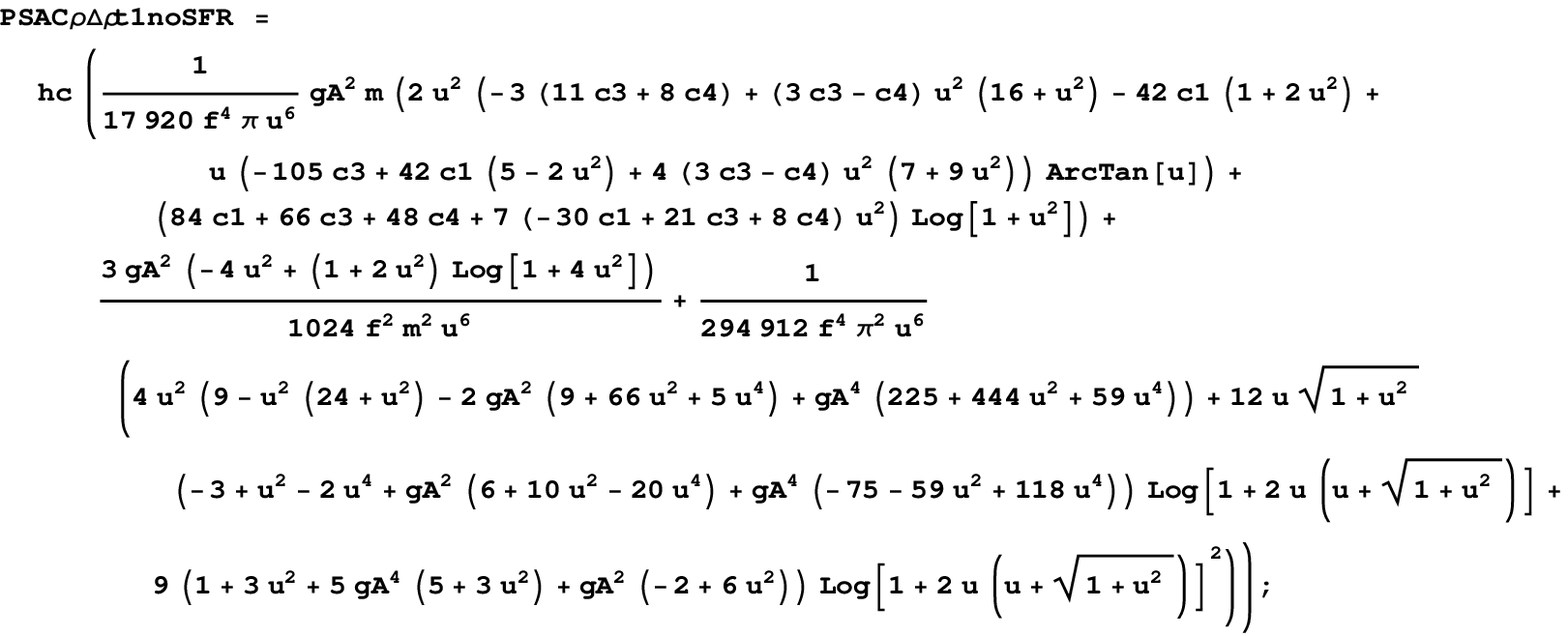}
\end{figure}

\begin{figure}[p]
\includegraphics{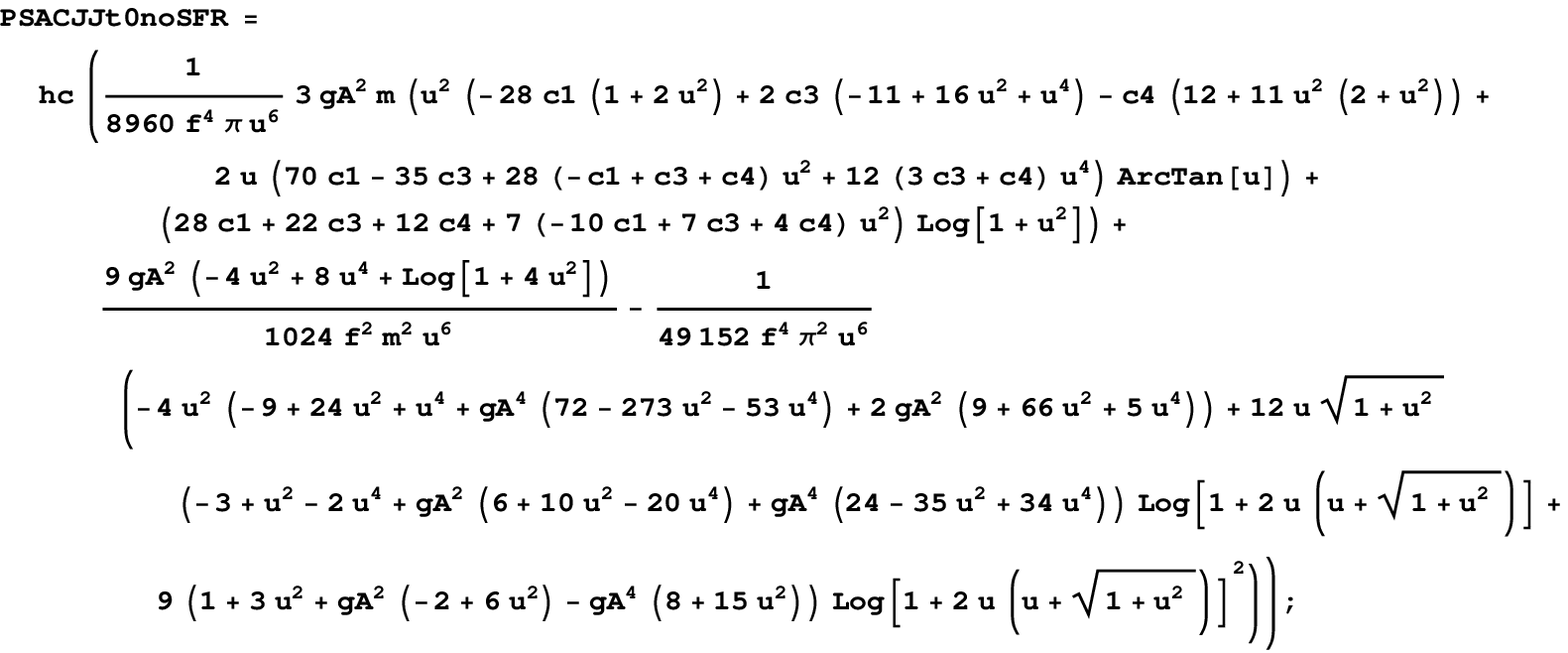}
%
\includegraphics{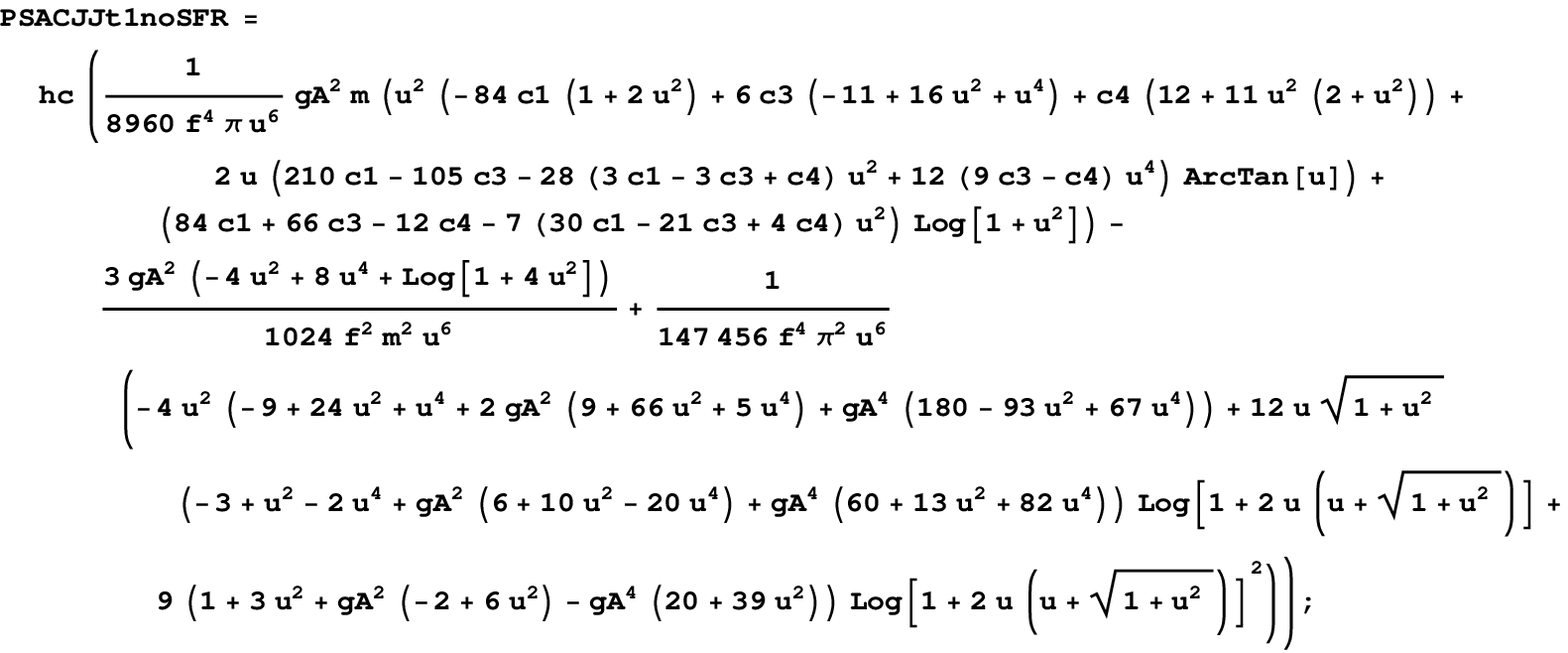}
\end{figure}
\end{widetext}
}
\afterpage{
}


\end{document}